\documentclass[10pt,a4paper]{article} 
\usepackage{graphicx}
\usepackage{epstopdf}
\usepackage{epsfig}
\usepackage{amsmath,amssymb,bm,amsfonts,bbm}
\usepackage{style}
\usepackage{yfonts}
\usepackage{subfig,caption}

\newcommand{\arXiv}[2]{\href{http://arxiv.org/pdf/#1}{{\tt [#2/#1]}}}

\newcommand{\be}{\begin{equation}}
\newcommand{\ee}{\end{equation}}
\newcommand{\bea}{\begin{eqnarray}}

\newcommand{\eea}{\end{eqnarray}}

\newcommand{\kx}{\kappa}
\newcommand{\mpl}{M_{\rm Pl}}
\newcommand{\Rb}{\bar{R}}

\def\Hc{{\cal H}}

\newcommand{\dphi}{\delta \varphi}

\def\exx{\varepsilon}

\def \gta {\mathrel{\vcenter
     {\hbox{$>$}\nointerlineskip\hbox{$\sim$}}}}

\DeclareSymbolFont{mathscrUC}{U}{rsfs}{m}{n}  
\DeclareSymbolFont{mathscrLC}{OT1}{pzc}{m}{n} 

\makeatletter

\DeclareRobustCommand*{\mathscr}[1]{\gdef\F@ntPrefix{mathscr@char@}%
  \@EachCharacter #1\@EndEachCharacter}
\long\def\DoLongFutureLet #1#2#3#4{%
   \def\@FutureLetDecide{#1#2\@FutureLetToken
      \def\@FutureLetNext{#3}\else
      \def\@FutureLetNext{#4}\fi\@FutureLetNext}
   \futurelet\@FutureLetToken\@FutureLetDecide}
\def\DoFutureLet #1#2#3#4{\DoLongFutureLet{#1}{#2}{#3}{#4}}
\def\@EachCharacter{\DoFutureLet{\ifx}{\@EndEachCharacter}%
   {\@EachCharacterDone}{\@PickUpTheCharacter}}
\def\m@keCharacter#1{\csname\F@ntPrefix#1\endcsname}
\def\@PickUpTheCharacter#1{\m@keCharacter{#1}\@EachCharacter}
\def\@EachCharacterDone \@EndEachCharacter{}

\DeclareMathSymbol{\mathscr@char@A}{\mathord}{mathscrUC}{`A}
\DeclareMathSymbol{\mathscr@char@B}{\mathord}{mathscrUC}{`B}
\DeclareMathSymbol{\mathscr@char@C}{\mathord}{mathscrUC}{`C}
\DeclareMathSymbol{\mathscr@char@D}{\mathord}{mathscrUC}{`D}
\DeclareMathSymbol{\mathscr@char@E}{\mathord}{mathscrUC}{`E}
\DeclareMathSymbol{\mathscr@char@F}{\mathord}{mathscrUC}{`F}
\DeclareMathSymbol{\mathscr@char@G}{\mathord}{mathscrUC}{`G}
\DeclareMathSymbol{\mathscr@char@H}{\mathord}{mathscrUC}{`H}
\DeclareMathSymbol{\mathscr@char@I}{\mathord}{mathscrUC}{`I}
\DeclareMathSymbol{\mathscr@char@J}{\mathord}{mathscrUC}{`J}
\DeclareMathSymbol{\mathscr@char@K}{\mathord}{mathscrUC}{`K}
\DeclareMathSymbol{\mathscr@char@L}{\mathord}{mathscrUC}{`L}
\DeclareMathSymbol{\mathscr@char@M}{\mathord}{mathscrUC}{`M}
\DeclareMathSymbol{\mathscr@char@N}{\mathord}{mathscrUC}{`N}
\DeclareMathSymbol{\mathscr@char@O}{\mathord}{mathscrUC}{`O}
\DeclareMathSymbol{\mathscr@char@P}{\mathord}{mathscrUC}{`P}
\DeclareMathSymbol{\mathscr@char@Q}{\mathord}{mathscrUC}{`Q}
\DeclareMathSymbol{\mathscr@char@R}{\mathord}{mathscrUC}{`R}
\DeclareMathSymbol{\mathscr@char@S}{\mathord}{mathscrUC}{`S}
\DeclareMathSymbol{\mathscr@char@T}{\mathord}{mathscrUC}{`T}
\DeclareMathSymbol{\mathscr@char@U}{\mathord}{mathscrUC}{`U}
\DeclareMathSymbol{\mathscr@char@V}{\mathord}{mathscrUC}{`V}
\DeclareMathSymbol{\mathscr@char@W}{\mathord}{mathscrUC}{`W}
\DeclareMathSymbol{\mathscr@char@X}{\mathord}{mathscrUC}{`X}
\DeclareMathSymbol{\mathscr@char@Y}{\mathord}{mathscrUC}{`Y}
\DeclareMathSymbol{\mathscr@char@Z}{\mathord}{mathscrUC}{`Z}
\DeclareMathSymbol{\mathscr@char@a}{\mathord}{mathscrLC}{`a}
\DeclareMathSymbol{\mathscr@char@b}{\mathord}{mathscrLC}{`b}
\DeclareMathSymbol{\mathscr@char@c}{\mathord}{mathscrLC}{`c}
\DeclareMathSymbol{\mathscr@char@d}{\mathord}{mathscrLC}{`d}
\DeclareMathSymbol{\mathscr@char@e}{\mathord}{mathscrLC}{`e}
\DeclareMathSymbol{\mathscr@char@f}{\mathord}{mathscrLC}{`f}
\DeclareMathSymbol{\mathscr@char@g}{\mathord}{mathscrLC}{`g}
\DeclareMathSymbol{\mathscr@char@h}{\mathord}{mathscrLC}{`h}
\DeclareMathSymbol{\mathscr@char@i}{\mathord}{mathscrLC}{`i}
\DeclareMathSymbol{\mathscr@char@j}{\mathord}{mathscrLC}{`j}
\DeclareMathSymbol{\mathscr@char@k}{\mathord}{mathscrLC}{`k}
\DeclareMathSymbol{\mathscr@char@l}{\mathord}{mathscrLC}{`l}
\DeclareMathSymbol{\mathscr@char@m}{\mathord}{mathscrLC}{`m}
\DeclareMathSymbol{\mathscr@char@n}{\mathord}{mathscrLC}{`n}
\DeclareMathSymbol{\mathscr@char@o}{\mathord}{mathscrLC}{`o}
\DeclareMathSymbol{\mathscr@char@p}{\mathord}{mathscrLC}{`p}
\DeclareMathSymbol{\mathscr@char@q}{\mathord}{mathscrLC}{`q}
\DeclareMathSymbol{\mathscr@char@r}{\mathord}{mathscrLC}{`r}
\DeclareMathSymbol{\mathscr@char@s}{\mathord}{mathscrLC}{`s}
\DeclareMathSymbol{\mathscr@char@t}{\mathord}{mathscrLC}{`t}
\DeclareMathSymbol{\mathscr@char@u}{\mathord}{mathscrLC}{`u}
\DeclareMathSymbol{\mathscr@char@v}{\mathord}{mathscrLC}{`v}
\DeclareMathSymbol{\mathscr@char@w}{\mathord}{mathscrLC}{`w}
\DeclareMathSymbol{\mathscr@char@x}{\mathord}{mathscrLC}{`x}
\DeclareMathSymbol{\mathscr@char@y}{\mathord}{mathscrLC}{`y}
\DeclareMathSymbol{\mathscr@char@z}{\mathord}{mathscrLC}{`z}

\makeatother

\relax

\title{Spectrum oscillations from features in the potential of single-field inflation
}

\author{I. Dalianis, G.P. Kodaxis, I.D. Stamou, N. Tetradis and A. Tsigkas-Kouvelis}

\affiliation{Department of Physics, University of Athens, University Campus, Zographou 157 84, Greece}

\emailAdd{idalianis@phys.uoa.gr, gekontax@phys.uoa.gr, joanstam@phys.uoa.gr, ntetrad@phys.uoa.gr, atsigkas@phys.uoa.gr}

\abstract{
We study single-field inflationary models with steep step-like features
in the potential that lead to the temporary violation of the slow-roll conditions
during the evolution of the inflaton.
These features enhance the power spectrum of the curvature perturbations 
by several orders of magnitude at certain scales and also 
produce prominent oscillatory patterns.
We study analytically and numerically the inflationary dynamics. We 
describe quantitatively the size of the enhancement, 
as well as the profile of the oscillations, which are 
shaped by the number and position of 
the features in the potential.
The induced tensor power spectrum inherits the distinctive oscillatory profile of 
the curvature spectrum and is potentially detectable 
by near-future space interferometers.
The enhancement of the power specrtum by step-like features, 
though significant, may  be insufficient to trigger the production 
of a sizeable number of primordial black holes
if radiation dominates the energy density of the early universe. 
However, it can result in sufficient 
black hole production if the universe is dominated by non-relativistic  matter. 
For the latter scenario, we find 
that deviations from 
the standard monochromatic profile of the mass spectrum of primordial black holes 
are possible because of the multiple-peak structure of the curvature power spectrum.
}

\begin{document}

\maketitle

\section{Introduction} \label{intro}

\subsection{Oscillations in the power spectrum} \label{oscillations}

Inflationary models that predict  deviations from scale invariance at small scales
have been attracting a lot of attention in recent years.
During the early evolution of the universe,
a strong enhancement of the spectrum of primordial scalar perturbations can trigger
the gravitational collapse and the formation of primordial black holes (PBHs), which
 may survive until today in significant numbers in order to be detectable
 \cite{pbh1, Hawking:1971ei, Carr:1974nx, Carr:1975qj}. 
  This possibility has been studied 
 in great detail during the last years. (For reviews with extensive lists of references, see
 \cite{Carr:2016drx,Sasaki:2018dmp,Carr:2020xqk, Green:2020jor}.)
In addition, a potentially observable 
stochastic background of gravitational waves (GWs) 
is generated through the coupling of scalar and tensor modes at second order
\cite{Matarrese:1992rp,Matarrese:1993zf,Matarrese:1997ay, Mollerach:2003nq,  Noh:2004bc, Carbone:2004iv, Nakamura:2004rm, Baumann:2007zm, Ananda:2006af, Assadullahi:2009nf}.
The induced tensors are suppressed by the small value of the scalar
perturbations at the CMB scales \cite{Aghanim:2018eyx}, but may
be sizeable if the primordial density perturbations are enhanced at small scales. 
In this way, the relic GW stochastic background may provide a 
direct probe of the very early cosmic history. 
The detection prospects of induced GWs  
open a new window to probe the inflationary dynamics at small scales, for which 
cosmic microwave background (CMB) observables lack sensitivity.

A primordial scalar spectrum with a strong enhancement can be realized in 
various setups, 
such as through inflationary potentials that contain a near-inflection point 
\cite{inflectionorig, Yokoyama:1998pt, Cheng:2016qzb, Garcia-Bellido:2017mdw, Ezquiaga:2017fvi, Germani:2017bcs, Motohashi:2017kbs, Gong:2017qlj,  Ballesteros:2017fsr, Hertzberg:2017dkh, Cheng:2018yyr, Cicoli:2018asa, Ozsoy:2018flq, Biagetti:2018pjj,  Dalianis:2018frf,
Gao:2018pvq, Tada:2019amh, Dalianis:2019asr, Atal:2019erb, Mahbub:2019uhl, Mishra:2019pzq, Ballesteros:2019hus,
Nanopoulos:2020nnh,Stamou:2021qdk, Liu:2020oqe}, 
multi-field inflation \cite{GarciaBellido:1996qt, Kawasaki:1997ju, Frampton:2010sw, Clesse:2015wea,  Kawasaki:2016pql, Inomata:2017okj, Espinosa:2017sgp, Inomata:2017vxo, Kawasaki:2019hvt, Palma:2020ejf, Fumagalli:2020adf, Braglia:2020eai, Aldabergenov:2020bpt},
modified gravity
\cite{Kannike:2017bxn, Pi:2017gih, Fu:2019ttf, Dalianis:2019vit, Cheong:2019vzl, Lin:2020goi}, 
curvaton models \cite{Kawasaki:2012wr, Kohri:2012yw, Ando:2017veq, Ando:2018nge}, 
sound speed modulation and parametric resonance 
\cite{Cai:2019bmk, Cai:2018tuh, Cai:2019jah, Chen:2019zza, Chen:2020uhe, Zhou:2020kkf}. 
It can also be realized when the inflationary potential features a step-like change \cite{Starobinsky:1992ts, Adams:2001vc, Leach:2000yw, Leach:2001zf, Hazra:2010ve}, 
a framework that was revisited recently in \cite{Kefala:2020xsx}.
It is very interesting that the enhancement profiles produced 
by these inflationary models may be distinguishable. 
Different inflationary realizations yield 
power spectra with a wide or narrow peak, oscillations or a multi-peak structure.

In this work we focus on power spectra with oscillations around the peak. 
The oscillatory pattern is distinctive and possibly detectable, indicating
a sharp feature in the inflationary dynamics. It can be caused by the reentry 
of $k$-modes in the horizon, a change in the sound speed 
\cite{Ballesteros:2018wlw},  
the backreaction of the entropy modes on the adiabatic modes 
in multi-field inflation \cite{Fumagalli:2020adf,Fumagalli:2020nvq, Braglia:2020taf,Fumagalli:2021cel},
or  by a step in the inflaton potential 
\cite{Starobinsky:1992ts, Adams:1997de,
Adams:2001vc, Leach:2000yw, Leach:2001zf, Covi:2006ci, Hamann:2007pa, Hazra:2010ve, Liu:2010dh, Cadavid:2015iya, GallegoCadavid:2016wcz, Fard:2017oex, Kefala:2020xsx}. 
The last example is the minimal realization of a sharp feature that 
involves single-field inflation dynamics and a canonical kinetic term. 
Motivated by the original proposal \cite{Starobinsky:1992ts}, 
where the effects of singular points in the inflationary potential were studied, 
we study here 
smooth variations of the basic setup, focusing on model-independent features.
We compute analytically and numerically the evolution of the curvature 
perturbations and find  a strong enhancement of the scalar spectrum. In addition,
we observe a burst of oscillations, generated solely by 
step-like changes in the inflaton potential.

As we show in the following section, a sharp drop in the 
potential of the inflaton field  detunes the relative phase between the 
real and imaginary parts of the curvature perturbation, so that 
oscillations in the amplitude  of the spectrum  appear, 
while no reentry of modes takes place.
The characteristic period of the oscillations depends on the position of 
the feature, while
interference patterns are also apparent. 
We demonstrate that, even though the strong features in the underlying inflaton 
evolution may not be simple and the range of generated spectra extensive, 
an analytical understanding of their form is feasible. 
This is achieved by approximating the time-dependent inflaton
background through a series of 
``pulses" that affect the evolution of the fluctuations.
A similar approach has been followed in ref. \cite{Kaloper:2003nv, DAmico:2020euu, DAmico:2021vka} in order to study
inflation that is realized through
a series of bursts of cosmic acceleration, separated by
intervals of decelerated expansion. Our setup can be viewed as a reduced version of  
the so-called ``rollercoaster cosmology" \cite{DAmico:2020euu}.

The amplitude of the peak of the spectrum of curvature perturbations 
${\cal P_R}(k)$ is determined by the characteristics of the
features in the inflaton potential, 
so that significant PBH production can be generated. 
It is exciting that the shape characteristics of the  
peak of the curvature spectrum 
can also be imprinted on the spectrum of induced GWs 
\cite{Fumagalli:2020adf, Fumagalli:2020nvq, Braglia:2020taf,Fumagalli:2021cel}. 
Specific realizations of this possibility involve
non-geodesic motion during multi-field inflation, 
or resonance effects. However, the link between strong features in the inflaton evolution and
strong oscillations in the curvature and GW spectra is generic,
as has been discussed in the above references.  

In general, the ${\cal P_R}(k)$ characteristics are not clearly visible in the 
mass spectrum of the fractional PBH abundance $f_\text{PBH}$, 
which appears predominantly monochromatic, 
mostly sensitive to the amplitude of the peak.
Even though a universal behavior also appears in the GW spectrum \cite{Cai:2019cdl},   
especially for smooth scalar spectra,  
the tensor perturbations are much more informative \cite{Pi:2020otn, Ananda:2006af,Saito:2008jc, Dalianis:2020cla} and can display more clearly
features originating in the scalar spectrum. In this way,
the detection of stochastic GWs is a portal to the primordial spectrum of 
scalar perturbations at small scales, which can also be used to test the 
PBH dark matter scenario. Moreover, it can provide details of the 
possible strong features in the inflationary dynamics.

The induced GWs may be detected in the near future by the current and planned 
detectors.
The LIGO collaboration has already produced upper limits in such 
stochastic backgrounds~\cite{TheLIGOScientific:2016dpb}. 
The searches will be further extended 
by a network of operating and designed gravitational wave detectors that will 
probe a vast range of different frequency bands.  Pulsar time array (PTA) GW 
experiments \cite{Chen:2019xse} have a sensitivity to the nano-Hz frequency band, 
space-based interferometers like LISA \cite{Audley:2017drz}, 
Taiji \cite{Guo:2018npi}, Tianqin \cite{Luo:2015ght}, 
Decigo \cite{Seto:2001qf, Sato:2017dkf} are mostly sensitive to milli-Hz
and deci-Hz frequency bands, and the LIGO/Virgo and  
Einstein telescope \cite{Sathyaprakash:2012jk} ground-based 
interferometers are sensitive to larger frequencies.

\subsection{The steps in the inflaton potential} \label{stepalpha}

In the following section we shall discuss in detail the oscillatory patterns
in the spectra of curvature perturbations and induced GWs that arise from 
steep steps in the inflaton potential. The steps connect regions in which
the potential varies smoothly and the slow-roll conditions are satisfied.
The basic pattern corresponds to the
vacuum energy having one or more transition points at which it jumps 
from one constant value to another \cite{Kefala:2020xsx}. 
One can speculate that these points may correspond to 
values of the inflaton field at which 
certain modes, whose quantum fluctuations contribute 
to the vacuum energy, decouple very quickly. 
Decoupling effects become visible when the effective 
potential is regularized in a mass-sensitive scheme. Also, the dependence of the
potential on an energy scale, or a coarse-graining length, can 
be analysed through the Wilsonian approach to the renormalization group,
see for example refs. \cite{Wetterich:1992yh, Berges:2000ew} for a particular
implementation. The resulting renormalization-group equation for the potential
can capture the decoupling behavior. 
However, our fundamental lack of understanding of the
nature of vacuum energy or the cosmological constant 
does not permit a quantitative calculation of these effects.

Some intuition on this issue can be obtained by considering 
the role of underlying symmetries. A specific framework, which we shall
use as the basis for the potentials that we shall consider, is provided by the 
models of $\alpha$-attractors in supergravity \cite{Kallosh:2013hoa, Ferrara}.
A toy model that demonstrates the role of symmetries 
is described by the Lagrangian \cite{Kallosh:2014rga} 
\be
{\cal L}=\sqrt{-g}\left[ 
\frac{1}{2}\partial_\mu \chi \partial^\mu \chi +\frac{1}{12}\chi^2 R(g)
-\frac{1}{2}\partial_\mu \phi \partial^\mu \phi -\frac{1}{12}\phi^2 R(g)
-\frac{1}{36}F^2(\phi/\chi) \left(\chi^2-\phi^2 \right)^2
\right],
\label{lagralpha} \ee
which is invariant under the conformal transformation 
\be
g_{\mu\nu}\to e^{-2\sigma(x)}g_{\mu\nu},~~~~~~\phi\to e^{\sigma(x)}\phi,~~~~~~\chi\to e^{\sigma(x)}\chi.
\label{conformal} \ee
For constant $F(\phi/\chi)$, there is a global 
$SO(1,1)$ symmetry that keeps $\chi^2-\phi^2$ constant. 
The field $\chi$ does not have any physical degrees of freedom and
can be eliminated through the gauge-fixing condition $\chi^2-\phi^2=6$. (All 
dimensionful quantities are expressed in units of $\mpl$.)
We parametrize the
fields as  
$\chi=\sqrt{6}\cosh(\varphi/\sqrt{6})$, $\phi=\sqrt{6}\sinh(\varphi/\sqrt{6})$
\cite{Kallosh:2014rga}.
The Lagrangian becomes
\be
{\cal L}=\sqrt{-g}\left[ 
 \frac{1}{2} R(g)-\frac{1}{2}\partial_\mu \varphi \partial^\mu \varphi
-F^2\left(\tanh \frac{\varphi}{\sqrt{6}} \right) \right].
\label{lagralphagauge} \ee
A constant function $F(x)$, which preserves the $SO(1,1)$ symmetry,
results in a cosmological constant in this formulation. 
The value of the cosmological constant is not constrained by the symmetry and is arbitrary.

We can introduce a minimal deformation of the symmetry by assuming that $F(x)$ 
takes two different values 
over two continuous ranges of $x$, with a rapid transition in between. 
A stronger deformation that has been used extensively in the literature 
assumes that $F(x)$ has a polynomial form.
We shall employ a combination of the above choices by assuming that 
$F(x)$ has the schematic form
\be
F(x)=x^n+ \sum_i A_i\, \Theta(x-x_i),
\label{fx} \ee
allowing for more than one transition points.
In order to avoid unphysical features in the evolution of the inflaton, 
each step-function is replaced by a continuous function with a sharp transition
at $x_i$. 
A more general framework is provided by the $\alpha$-attractors \cite{Kallosh:2013hoa, Ferrara, Kallosh:2014rga}. The 
Lagrangian includes an additional free parameter $\alpha$ and takes the form 
\be
{\cal L}=\sqrt{-g}\left[ 
 \frac{1}{2} R(g)-\frac{1}{2}\partial_\mu \varphi \partial^\mu \varphi
-F^2\left(\tanh \frac{\varphi}{\sqrt{6\alpha}} \right) \right].
\label{lalph} \ee

The potentials that result from our assumption for the function $F(x)$ 
with positive $A_i$ are
generalizations of the potential of the Starobinsky model 
 \cite{Starobinsky:1980te}, with the addition of one or more steep steps. 
Allowing for negative values of $A_i$ makes it possible to include inflection points
 in the potential as well. As our analysis focuses on the phenomenological consequences of 
 general features in the potential,
 we consider parameters $A_i$ that can take values over the whole real axis.
 Another important feature of the potential in eq. (\ref{lalph}) is the sharpness
 of the transition between ranges of constant vacuum energy. This transition is modelled by a 
 $\Theta$-function in eq. (\ref{fx}), but it is smooth in practice. Its steepness
 affects the oscillatory patterns appearing in the spectra. Because of our lack of understanding
 of the essence of the cosmological constant, we refrain from explicit model-building, and
 treat the steepness as a free parameter. We only point out that the framework of 
 $\alpha$-attractors results in the dependence of the potential on
 $\tanh ({\varphi}/\sqrt{6\alpha})$, with $\alpha$ a free parameter. This allows, in principle, for
 potentials with transitions of arbitrary steepness.

\subsection{Plan of the paper} \label{plan}

The plan of the paper is as follows:
In the next section we present an analytical discussion of the oscillatory 
patterns that can appear in the spectrum of curvature perturbations 
when the inflaton potential contains step-like features. We first establish
our notation and identify the relevant parameters for the analysis of 
oscillations. We next use a 
simple toy model, neglecting the expansion of the background, in order to 
demonstrate how the detuning of the relative phase between the real and imaginary
parts of the perturbation generates an oscillatory pattern in its amplitude.
We then present an analytical study of the oscillatory 
form of the curvature power spectra that 
may result from strong features in the inflaton potential. This is possible
if the effect on the fluctuations is modelled by a series of positive or 
negative ``pulses"  that correspond to the deviations from the slow-roll regime.
In section \ref{blackholesGW} we study explicit inflationary realizations with 
step-like features in the framework of $\alpha$-attractors, 
paying particular attention to the consistency with the CMB constraints. 
For these inflationary models we examine the production of PBHs and their mass distribution, 
as well as the spectrum of the induced GWs.
We elaborate on the relations and similarities  between the patterns appearing in the  
curvature and tensor power spectra.
The last section \ref{conclusions} contains our conclusions.
All dimensionful quantities are given in units of $\mpl$ throughout the paper, unless the
units are explicitly stated. 

\section{Analytical calculation of the spectrum of curvature perturbations} \label{notes}

\subsection{General considerations} \label{general}

In this section we discuss an approximate analytical treatment of the spectrum of curvature 
perturbations in cases that the slow-roll approximation is strongly violated. 
We assume that the inflaton potential displays the standard plateau that can lead to 
an almost scale-invariant spectrum. In addition, it contains a strong feature within a finite range of 
field values, which can lead to the violation of the slow-roll conditions or even cause inflation
to cease momentarily. In order to be as model independent as possible, we do not focus on 
specific potentials with these properties.

We consider the most general scalar metric perturbation 
around the Friedmann-Robertson-Walker 
(FRW) background \cite{Mukhanov:1990me}
 \be
 ds^2=a^2(\tau)\left\{ (1+2\phi)d\tau^2-2B_{,i}\,dx^i d\tau
 -\left((1-2\psi)\delta_{ij}+2E_{,ij} \right)dx^i dx^j
 \right\},
 \label{metric} \ee
 with $B_{,i}=\partial_i B$, $E_{,ij}=\partial_i\partial_jE$.
The inflaton field can be split into a background and a perturbation:
$\varphi(\tau)+\delta\varphi(\tau,x)$. A gauge-invariant field perturbation can be defined as
$
 v =a\left( \dphi+(\varphi'/\Hc)\psi \right)$,
satisfying the Mukhanov-Sasaki equation \cite{Mukhanov:1988jd, Sasaki:1986hm}
\be
v''-\nabla^2 v-\frac{z''}{z}\, v=0,
\label{eomv} \ee
with $z=a\varphi'/\Hc$. The primes and the Hubble parameter 
refer to derivatives with respect to conformal time.
The 
gauge-invariant comoving curvature perturbation $R=-v/z$ satisfies 
\be
R_k''+2\frac{z'}{z}R_k'+k^2 R_k=0
\label{geqqf} \ee
in Fourier space.

We shall use the number of
efoldings $N$ as the independent variable for the evolution of the
perturbations. 
The Hamilton-Jacobi slow-roll parameters are defined through the relations
\begin{eqnarray}
H^2&=&\frac{V(\varphi)}{3\mpl^2-\frac{1}{2}\varphi_{,N}^2}
\label{HH22} \\
\exx_H&=&
-\frac{d\ln H}{dN}
=\frac{ \varphi^2_{,N}}{2 \mpl^2}
\label{b1} \\
\eta_H&=&\exx_H-\frac{1}{2}\frac{d \ln \exx_H}{dN}=
\frac{ \varphi^2_{,N}}{2 \mpl^2}-\frac{\varphi_{,NN}}{\varphi_{,N}},
\label{b2} \end{eqnarray}
where $H=e^{-N}\Hc$ is the Hubble parameter defined through cosmic time,
and subscipts denote derivatives with respect to $N$.
The parameter $z$ is defined as 
$z= e^N\, \varphi_{,N}.$
The effective equation of state for the background is 
$w=-1+2\exx_H/3$.
The equation for the curvature perturbation takes the form
\be
R_{k,NN}+f(N)\, R_{k,N}+\frac{k^2}{e^{2N} H^2} R_{k}=0,
\label{RN} \ee
with the quantity 
\be
f(N)=3+\frac{2\varphi_{,NN}}{\varphi_{,N}}-\frac{\varphi_{,N}^2}{2\mpl^2}=3+\exx_H-2\eta_H
\label{AN} \ee
playing a crucial role in determining the qualitative 
behaviour of the solutions.
In the slow-roll regime
it acts as a generalized friction term. However, if $\eta_H$ becomes 
positive and large it can lead to
a dramatic enhancement of the perturbations. 

In the approximation that the slow-roll parameters are neglected and $H$ is assumed to remain 
constant, the solution of eq. (\ref{RN}) can be expressed in terms of the 
Bessel functions  $J_{\pm 3/2}$ as
\be
R_k(N;C_{p},C_{m},3)=A \, e^{-\frac{3}{2}\, N} 
\left(C_{p}J_{3/2}\left( e^{- N}\frac{k}{H} \right) + C_{m}\, J_{-3/2}
\left( e^{-N}\frac{k}{H} \right)  \right),
\label{solBessel3} \ee
where we take $A$ to be real without loss of generality.
For the values $C_p=1$, $C_m=i$ the two Bessel functions combine into the
Hankel function of the first kind $H^{(1)}_{3/2}$. The curvature pertubation is
$R_k(N;1,i,3)\propto \left( e^{-i k \tau}/\sqrt{k} \right) \, (1-i/(k\tau))/ a(\tau)$, 
where the
conformal time is $H \tau=-e^{-N}=-1/a$. For $\tau \to -\infty$ this
is the standard expression for the Bunch-Davies vacuum in the slow-roll regime, which 
is taken as the initial condition for the evolution of the fluctuations. 
For $\tau \to 0^-$ the curvature perturbation approaches a constant value $\propto k^{-3/2}$ as the 
mode with wavenumber $k$ moves out of the horizon and freezes.
The power spectrum of curvature perturbations $\Delta_R^2=({k^3}/{2\pi^2})|R_k|^2$ is scale invariant.
It is important to notice that the value of the curvature perturbation at late times, or $N\to \infty$,
comes from the second term in eq. (\ref{solBessel3}), as the first one vanishes. In this sense, it
is the absolute value of $C_m$ that determines the power spectrum. 

The above picture is modified when the function $f(N)$ of eq. (\ref{AN}) deviates from a constant value
equal to 3. For small values of $\exx_H$, $\eta_H$ the deviations from scale invariance 
can be computed analytically through the standard slow-roll analysis. However, our
interest lies with strong modifications of $\exx_H$, $\eta_H$ that result in the enhancement of
the spectrum by several orders of magnitude.

\begin{figure}[t!]
\centering
\includegraphics[width=0.3\textwidth]{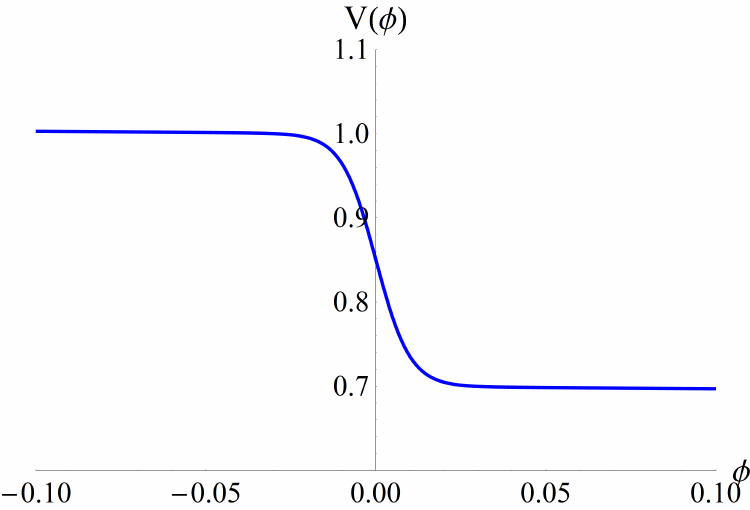}
\hspace{0.2cm} 
\includegraphics[width=0.3\textwidth]{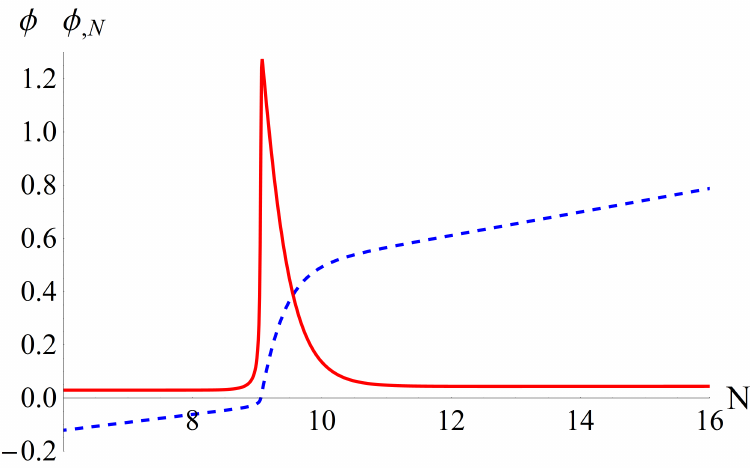}
\hspace{0.2cm} 
\includegraphics[width=0.3\textwidth]{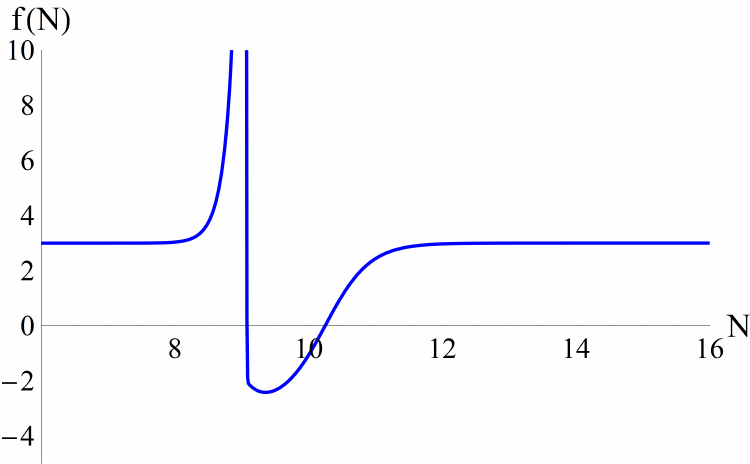} 
\\
\vspace{0.5cm}
\includegraphics[width=0.30\textwidth]{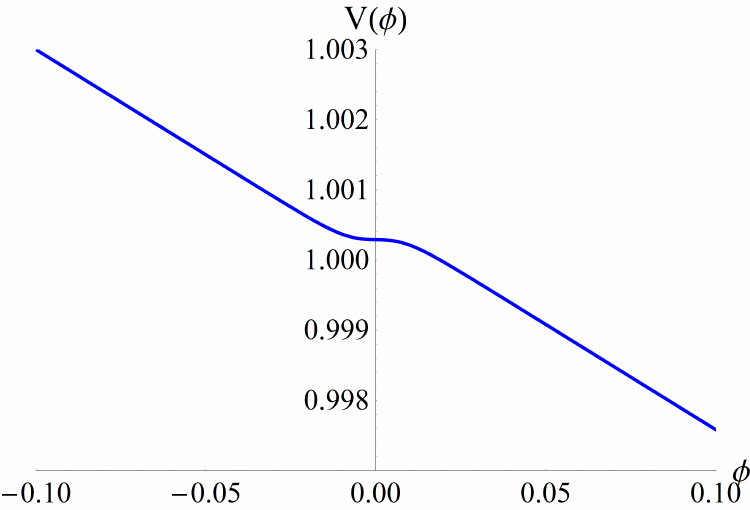}
\hspace{0.2cm} 
\includegraphics[width=0.30\textwidth]{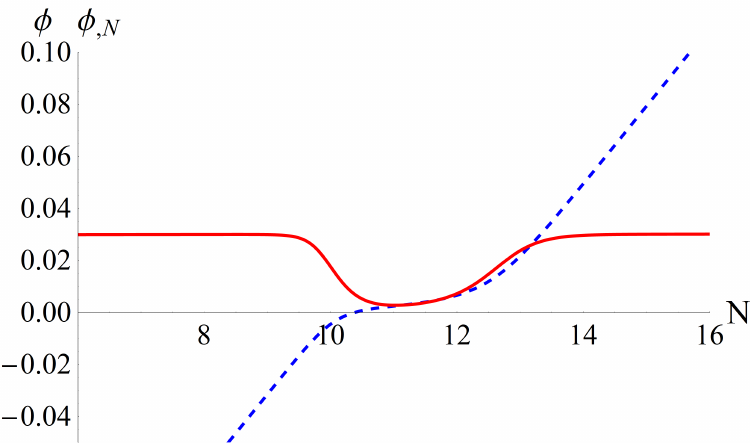}
\hspace{0.2cm} 
\includegraphics[width=0.30\textwidth]{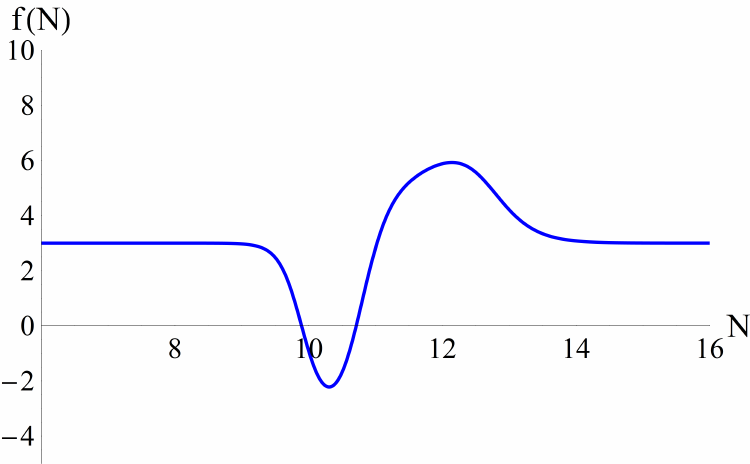} 
\caption{
The inflaton potential $V(\varphi)$ of eq. (\ref{pote}), the evolution of the inflaton
$\varphi$ (dashed line) and its derivative $\varphi_{,N}$ (solid line), 
and the function
$f(N)$ defined in eq. (\ref{AN}),
for two choices of the parameters of the potential: 
First row: $C=-0.3$, $c=100$, $B=-0.03$.~
Second row: $C=0.00058$, $c=100$, $B=-0.03$.
}
\label{friction}
\end{figure}

The typical forms of the effective-friction function $f(N)$ 
that we would like to analyse are depicted in
fig. \ref{friction}. 
These examples result from an inflaton potential used in ref. \cite{Kefala:2020xsx},
\be
V(\varphi)=V_0\left( 1+\frac{1}{2} C 
\left( 1+\tanh(c \varphi) \right)+B\varphi \right),
\label{pote} \ee
for specific choices of its parameters.
The function $f(N)$ remains close to 3, apart from a range of efoldings in which 
it deviates strongly from this value.
Similar features can be obtained with other types of 
potentials in single- or multi-field inflation, such 
as potentials with inflection points, or multiple inflationary stages. 
 The pattern can be repeated several times. 
When $f(N)$ becomes negative it induces a strong 
enhancement of the spectrum. The modifications to the spectrum
appear for wavenumbers of density perturbations
deep in the nonlinear regime today.

In order to obtain an analytical solution, 
we model $f(N)$ through a sequence of square ``pulses", each with constant $f(N)=\kappa_i\not=3$. 
At early and late times we assume that the inflaton is in a slow-roll regime, with negligible 
slow-roll parameters, so that $f(N)=3$ and the curvature perturbation is given by eq. (\ref{solBessel3}).
We approximate the Hubble parameter $H$ as almost constant. This is a good approximation, as our focus
is on modifications of the spectrum by several orders of magnitude. In comparison, the change in
the Hubble parameter for an
inflection point in the potential is less than 1\%, 
while for a step in the potential 
it is of order 10\%. 
We use an arbitrary normalization for the number of efoldings by absorbing a factor of 
$\exp(N_0)$ in $k$, where $N_0$ corresponds to the actual number of efoldings since the 
beginning of inflation until the moment in time that we denote by $N=0$. 
In practice this means that the physical value of the wavenumber is $\exp(N_0)\, k$.

Our starting point is the solution (\ref{solBessel3}), which defines the initial condition
for $N\to -\infty$. For $C_p=1$, $C_m=i$, this expression corresponds to the Bunch-Davies vacuum.
We neglect slow-roll corrections and approximate the evolution through eq. (\ref{solBessel3}) until
the value of $N$ at which the first nontrivial ``pulse" appears in $f(N)$. In the following 
subsection we analyse the modification of the curvature perturbation induced by this and the 
following ``pulses'', until the system returns to a slow-roll regime. 
For $N \to \infty$ the solution becomes constant. 
We are interested in the relative increase of the asymptotic value of $|C_m|$ in comparison to the 
value $|C_m|=1$ corresponding to a scale-invariant spectrum. 
In this sense, the value of the $k$-independent parameter $A$ in eq. (\ref{solBessel3}) is not
of interest to us. This parameter would determine the amplitude of the spectrum in the CMB region, 
and needs to be adjusted to a phenomenologically correct value. 

An important point concerns the form of $f(N)$. Negative values of 
this function result only from $\eta_H$ taking large positive values, as can be 
seen through eqs. (\ref{b1}), (\ref{b2}), (\ref{AN}). 
In general, large deviations from 3 can result from 
the term $2 \varphi_{,NN}/\varphi_{,N}$ being the dominant one in eq. (\ref{AN}).
The integral of $f(N)-3$ over $N$, from an early to a late slow-roll regime separated
by nontrivial evolution, is
\be
\int_{N_e}^{N_l} dN (f(N)-3)=
\int_{N_e}^{N_l} dN (\exx_H-2\, \eta_H)=2\log\frac{\left(\varphi_{,N}\right)_l}{\left(\varphi_{,N}\right)_e}+ \log\frac{H_l}{H_e}
=\log\frac{\left(dH/dN\right)_l}{\left(dH/dN\right)_e},
\label{integra} \ee 
where we have used the definitions (\ref{b1}), (\ref{b2}).
This quantity is approximately zero 
for inflaton potentials with a strong feature localized within 
a region supporting slow-roll inflation and with similar values of $dH/dN$ before and after the feature.
In this work we neglect the slow-roll 
corrections and analyse only the very large enhancement resulting from such a 
strong feature, by imposing the constraint 
that positive and negative ``pulses" have integrated 
areas that cancel.

\subsection{Toy-model analysis} \label{toy}

Several features that appear in the spectra that we study in the following sections
can be understood in a much simpler context. We are interested in the effect of 
a ``pulse" on the evolution of a mode with a free-wave initial condition. 
It is instructive to ignore the background expansion and consider the 
toy-model equation 
\be
R_{k,tt}+\kx R_{k,t}+ k^2 R_{k}=0.
\label{wave} \ee
The solutions are oscillatory with an amplitude that gets suppressed or enhanced,
depending on the sign of the friction parameter $\kx$.
It is straightforward to derive the solution for a friction term that vanishes
at all times apart from the interval $0<t<t_p$, by requiring the continuity of the 
solution and its first derivative at $t=0$ and $t_p$.

\begin{figure}[t!]
\centering
\includegraphics[width=0.4\textwidth]{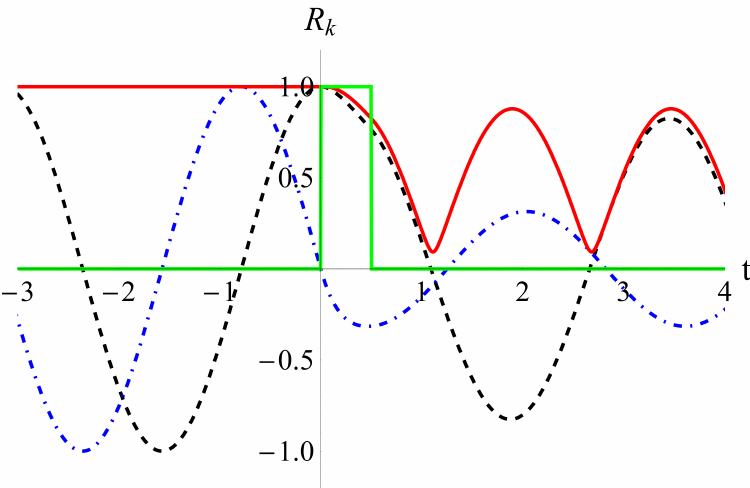}
\hspace{0.5cm}
 \includegraphics[width=0.4\textwidth]{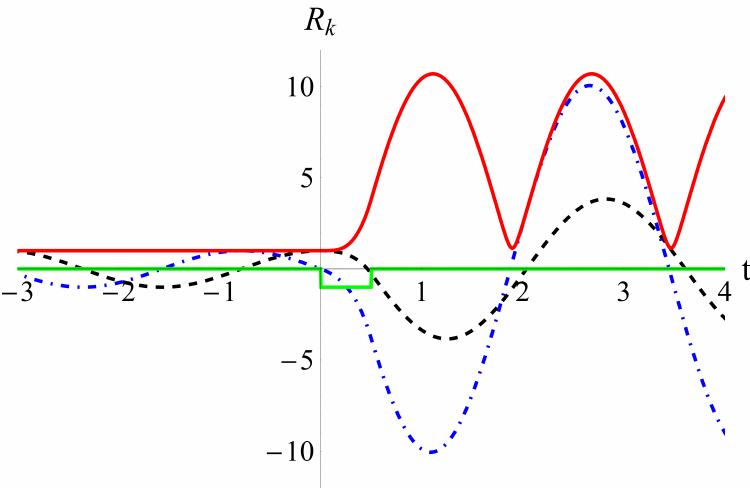} 
\caption{
The real part (dashed curve), imaginary part (dot-dashed curve) and 
amplitude (solid curve) of the solution of eq. (\ref{wave}) with $k=2$, for a ``pulse" in 
the interval $0\leq t \leq 0.5$. We also display the ``pulse'', with a rescaled
maximum $\kx/5$.
Left plot: $\kx=5$. Right plot: $\kx=-5$. 
}
\label{plotwave1}
\end{figure}

For an early-time solution
$R_k(t)=e^{-i k t}$, the evolution is depicted in fig. \ref{plotwave1}.
We observe the suppression of the amplitude for positive $\kx$ and the enhancement for
negative $\kx$. However, the most striking feature is the appearance of 
oscillations in the amplitude. Their origin lies in the modification by the ``pulse" of
the relative phase between the real and imaginary parts. 
For sufficiently large $|\kx|$ the relative phase in the
late stage of the evolution almost vanishes (as in the plot), so
that the amplitude approaches zero at certain instances.
In the cosmological context, 
the oscillatory form of the evolution as a function of time
can be transferred to the spectrum of perturbations. At late times, each
mode $k$ exits the horizon and eventually freezes. This can occur at any 
point of the oscillatory cycle, depending on the value of $k$.
As a result, the asymptotic values of the perturbations depend strongly on the
freezing time, and the
spectrum displays oscillations as a function of $k$.

It is known that it is possible to obtain an oscillatory pattern in the spectrum if
inflation stops for a certain time interval, so that modes that had exited 
the horizon reenter and start oscillating again until their next exit. Our toy example implies 
a more general pattern: Any feature during the evolution of the perturbations that
detunes the relative phase between the real and imaginary parts of the solution
results in an oscillatory spectrum, even if inflation is not halted.

\begin{figure}[t!]
\centering
\includegraphics[width=0.4\textwidth]{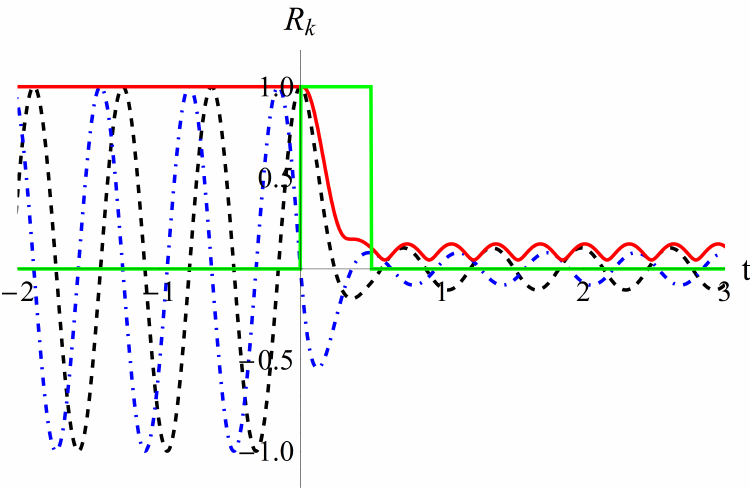}
\hspace{0.5cm}
 \includegraphics[width=0.4\textwidth]{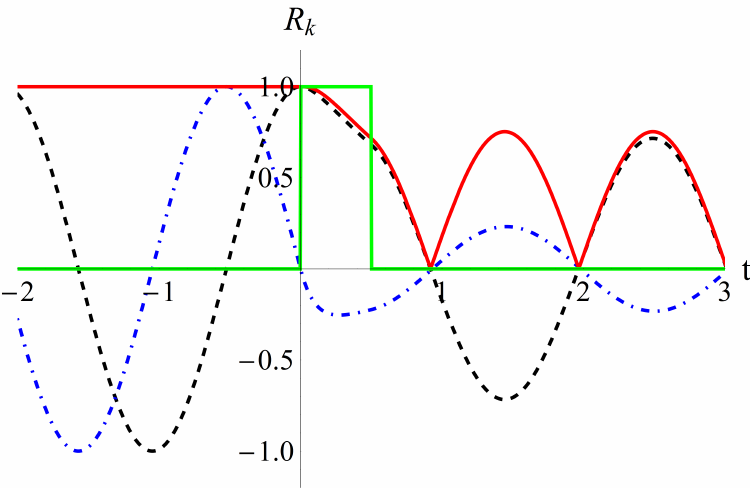} 
\caption{
The real part (dashed curve), imaginary part (dot-dashed curve) and 
amplitude (solid curve) of the solution of eq. (\ref{wave}) for a ``pulse" 
with $\kx=10$ in 
the interval $0\leq t \leq 0.5$.
We also display the ``pulse'', with a rescaled
maximum $\kx/10$.
Left plot: $k=10$; right plot: $k=3$, in arbitrary units. 
}
\label{plotwave2}
\end{figure}

Another interesting property of the late-time evolution 
is displayed in fig. \ref{plotwave2}: The relative suppression of the amplitude of 
a mode for a given positive friction parameter $\kx$ is larger for higher wavenumber $k$.
This is counter-intuitive at first sight, as one would expect the last term 
of eq. (\ref{wave}) to become more dominant for larger $k$ and limit the suppression
induced by the second term.
However, the opposite happens. For small $k$, the strong friction 
tends to freeze the evolution during the ``pulse", so that  
the real and imaginary parts resume their oscillations
after the ``pulse" with amplitudes comparable to the initial ones. 
As a general rule of thumb, for a duration of the ``pulse" of order 1, 
a strong suppression of the solution occurs for $k \gta \kx$.

\subsection{Analytical expressions for ``pulses"} \label{analytexpr}

\begin{figure}[t!]
\centering
\includegraphics[width=0.45\textwidth]{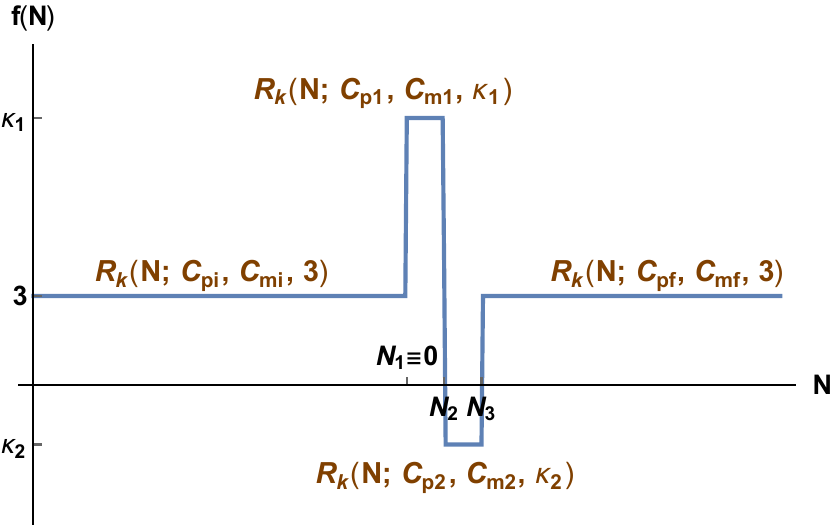}
\hspace{0.5cm}
 \includegraphics[width=0.46\textwidth]{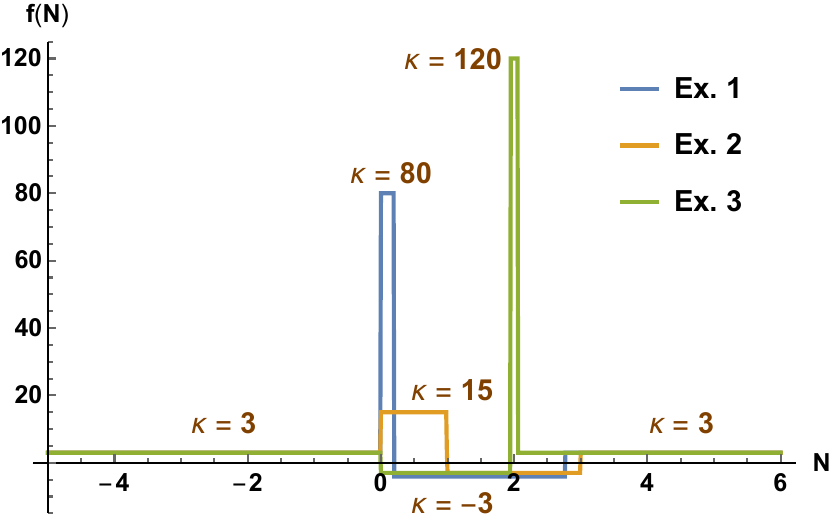} 
\caption{ An illustration of the approximate form 
of the function $f(N)$ that we assume for the analytical study. 
Left panel:  a double-``pulse" model, with a positive-friction ``pulse" followed by a negative-friction one.
Right panel: the double-``pulse" form assumed in the three examples resulting in the spectra 
of figs. \ref{spectrum1}, \ref{spectrum2}, \ref{spectrum3}.
}
\label{cartoon}
\end{figure}

We turn next to the analysis of eq. (\ref{RN}).
For constant $f(N)=\kappa$ the solution involves a linear combination of the Bessel functions  $J_{\pm\kappa/2}$ and has the form
\be
R_k(N;C_{p},C_{m},\kappa)=A e^{-\frac{1}{2}\kappa\, N} \left(C_{p}J_{\kappa/2}\left( e^{- N}\frac{k}{H} \right) + C_{m}\, J_{-\kappa/2}\left( e^{-N}\frac{k}{H} \right)  \right).
\label{solBessel}
 \ee
Let us suppose that the coefficients of the solution $C_{p_{i}}, C_{m_{i}}$ are known for a
range of efoldings for which 
$\kappa$ takes a specific value $\kappa_{i} $.
If this range is followed by a transition at $N=N_{fi}$ to a second range in which $\kappa$ takes a 
differrent value
$\kappa_{f}$, we would like to compute the corresponding values of the 
constants $ C_{p_{f}}, C_{m_{f}}$, see fig. \ref{cartoon}. 
This can be achieved by requiring the continuity of the solution and
its first derivative at $N=N_{fi}.$
A similar analysis has been performed in refs. \cite{Byrnes:2018txb,Carrilho:2019oqg}, using
conformal time as the independent variable.
We aim here at providing a more transparent picture of the oscillatory patterns in the 
spectrum, by identifying the characteristic frequencies. Moreover, in subection 
we provide an analytical treatment that goes beyond the modelling of $f(N)$ through 
square ``pulses".

The new coefficients  are given through the relation
\be
\begin{pmatrix}  
		 C_{p_{f}}\\
		  C_{m_{f}}
\end{pmatrix}=M(N_{fi},\kappa_{i},\kappa_{f},k)
\begin{pmatrix}	  
		 C_{p_{i}}\\
		  C_{m_{i}}
\end{pmatrix},
\label{coefficients}	  
\ee
where the matrix $M(N_{fi},\kappa_{i},\kappa_{f},k)$
has components
\begin{eqnarray}
	M_{11}&=&C \left( J_{-{\kappa_{f}}/{2}}\left(e^{-N_{fi}}\frac{k}{H}\right)
	J_{-1+\kappa_{i}/2}\left(e^{-N_{fi}}\frac{k}{H}\right)
	+J_{1-{\kappa_{f}}/{2}}\left(e^{-N_{fi}}\frac{k}{H}\right)
	J_{{\kappa_{i}}/{2}}\left(e^{-N_{fi}}\frac{k}{H}\right)\right)
	\nonumber \\ 
	M_{12}&=&C \left( -J_{-{\kappa_{f}}/{2}}\left(e^{-N_{fi}}\frac{k}{H}\right)
	J_{1-{\kappa_{i}}/{2}}\left(e^{-N_{fi}}\frac{k}{H}\right)
	+J_{1-{\kappa_{f}}/{2}}\left(e^{-N_{fi}}\frac{k}{H}\right)
	J_{-{\kappa_{i}}/{2}}\left(e^{-N_{fi}}\frac{k}{H}\right)\right)
	\nonumber \\ 
	M_{21}&=&C\left( -J_{{\kappa_{f}}/{2}}\left(e^{-N_{fi}}\frac{k}{H}\right)
	J_{-1+\kappa_{i}/2}\left(e^{-N_{fi}}\frac{k}{H}\right)
	+J_{-1+\kappa_{f}/2}\left(e^{-N_{fi}}\frac{k}{H}\right)
	J_{{\kappa_{i}}/{2}}\left(e^{-N_{fi}}\frac{k}{H}\right)\right)
	\nonumber \\ 
	M_{22}&=&C\left( J_{{\kappa_{f}}/{2}}\left(e^{-N_{fi}}\frac{k}{H}\right)
	J_{1-{\kappa_{i}}/{2}}\left(e^{-N_{fi}}\frac{k}{H}\right)
	+J_{-1+\kappa_{f}/2}\left(e^{-N_{fi}}\frac{k}{H}\right)
	J_{-{\kappa_{i}}/{2}}\left(e^{-N_{fi}}\frac{k}{H}\right)\right),
	\nonumber \\ &&
\label{Mab} \end{eqnarray}
with
\be
C=\frac{\pi}{2}e^{\frac{1}{2}N_{fi}(-2+\kappa_{f}-\kappa_{i})}\frac{k}{H} \,
\csc\left(\frac{\pi \kappa_{f}}{2}\right).
\label{Mconst} \ee
The matrix has the property $ M(N_{fi},\kappa_{m},\kappa_{f},k)\cdot M(N_{fi},\kappa_{i},\kappa_{m},k)=M(N_{fi},\kappa_{i},\kappa_{f},k)$.
This implies that we can select the value 
$\kappa=3$ as a reference point for all transitions between
different values of $\kappa$.

The next step is to define a matrix corresponding to a ``pulse" of height $\kappa$ above the 
value corresponding to the scale-invariant case. This matrix can be defined as
\be
M_{pulse}(N_1,N_2,\kappa,k)   =   M(N_{2},\kappa,3,k) \cdot M(N_{1},3,\kappa,k).
\label{Mpulse} \ee
As we explained earlier, the increase of the power spectrum relative to the 
scale invariant one is given by the value of $|C_m|^2$ after a mode of given $k$
has evolved past the strong features in the background. 
A product of several $M_{pulse}$ 
matrices can reproduce the final values of the coefficients 
$(C_p,C_m)$ of the Bessel functions $J_{\pm 3/2}$ after the fluctuations have evolved
from an initial configuration corresponding to $(C_p,C_m)=(1,i)$ through a period 
of strong features in the function $f(N)$. Clearly, it is possible to reconstruct
any smooth function $f(N)$ in terms of 
short intervals of $N$ during which the function takes constant values. Multiplying the
corresponding $M_{pulse}$ matrices would provide a solution to the problem of the 
evolution of perturbations. However, such an approach is not very efficient for 
a numerical solution.
We are mainly interested in obtaining intuitive analytical expressions
for forms of $f(N)$ such as those
depicted in fig. \ref{friction} and  \ref{cartoon}, for which a product of a small number of 
$M_{pulse}$ matrices suffices.

Simple analytical expressions can be obtained in the limits of large and small $k$,
using the corresponding expansions of the Bessel functions. 
For a large real argument we have
\be
J_{a}(z)=\sqrt{\frac{2}{\pi z}}\left[\cos\left(z-\frac{a\pi}{2}-\frac{\pi}{4}\right)-\frac{4a^{2}-1}{8z}\sin
\left(z-\frac{a\pi}{2}-\frac{\pi}{4}\right)+\mathcal{O}\left(z^{-2}\right)\right].
\label{expansion} \ee
Using this expression we find for large $k$ 
\begin{equation}
M_{pulse}^{(\infty)}(N_{1},N_{2},\kappa,k)=e^{-\frac{1}{2}(N_{2}-N_{1})(\kappa-3)}
\left\lbrace\begin{pmatrix} 1&0\\ 0&1 \end{pmatrix}
+\frac{1}{8}(\kappa-3) \frac{H}{k}\begin{pmatrix} S_{11}&S_{12}\\S_{21}&S_{22}
 \end{pmatrix}\right\rbrace
\label{Mpulseinf}
\end{equation}
where
\begin{eqnarray}
S_{11}&=&2 e^{N_{1}}\sin\left(2 e^{-N_{1}}\frac{k}{H}\right)-2 e^{N_{2}}\sin\left(2 e^{-N_{2}}\frac{k}{H}\right), \nonumber \\
S_{12}&=&-e^{N_{1}}\left(1+\kappa+2\cos\left(2 e^{-N_{1}}\frac{k}{H}\right)\right)+ e^{N_{2}}\left(1+\kappa+2\cos\left(2 e^{-N_{2}}\frac{k}{H}\right)\right),
\nonumber \\
S_{21}&=&e^{N_{1}}\left(1+\kappa-2\cos\left(2 e^{-N_{1}}\frac{k}{H}\right)\right)- e^{N_{2}}\left(1+\kappa-2\cos\left(2 e^{-N_{2}}\frac{k}{H}\right)\right),
\nonumber \\
S_{22}&=&-2 e^{N_{1}}\sin\left(2 e^{-N_{1}}\frac{k}{H}\right)+2 e^{N_{2}}\sin\left(2 e^{-N_{2}}\frac{k}{H}\right).
\label{Mpulseasympt} \end{eqnarray}
Keeping the leading contribution, we find that
the power spectrum is scale invariant  at late times (or $N\to \infty$) for $ k\rightarrow\infty$,  but has a value multiplied
by the factor
\be \label{enhancement}
\left[ \delta \Delta^{(\infty)}_{R}\right]^2=|C_m|^2=e^{-(N_{2}-N_{1})(\kappa-3)},
\ee 
relative to its scale-invariant value for modes that have sufficiently small $k$, so that
they exit the horizon and decouple very early with $C_m=i$, without being affected by the features in $f(N)$.
The exponent in the above expression is simply the area of the ``pulse" exceeding the value 3. 
For $\kappa >3$ the spectrum is suppressed, while for $\kappa <3$ it is enhanced. 
By breaking a general function $f(N)$ in infinitesimal ``pulses", it is easy to see
that the enhancement is equal to the integral of $f(N)-3$ over $N$.
The corrections subleading in $H/k$ introduce oscillatory patterns in the spectrum. 
The characteristic periods can be deduced from eqs. (\ref{Mpulseasympt}). 
The spectrum is expected to vanish at intervals $\delta k/H=e^{N_1}\pi$ and 
$\delta k/H=e^{N_2}\pi$. Moreover, when $N_1\simeq N_2$ we expect interference patterns.

Analytical expressions for $k\to 0$ are more difficult to obtain because
the (1,2)-component of the matrix $M_{pulse}$ scales as $1/k$ in this limit. As a result,
the effect of several ``pulses'', which involves the product of several such matrices,
is not described by a simple analytical expression. 
However, the components (2,1) and (2,2), which are relevant for the spectrum, are
simpler. The $(2,1)$-component becomes nonzero only at order $(k/H)^3$, while the 
$(2,2)$-component is equal to $1+{\cal O}\left( (k/H)^2 \right)$.
So, up to order $(k/H)^2$, the (2,2)-component is sufficient for the 
calculation of the spectrum. We give the result for the sequence of two ``pulses":
\begin{equation} \label{Mpulse0}
\begin{split}
M_{pulse}^{(0)}(N_{1},N_{2},N_3,\kappa_1,&\kappa_2,k)|_{2,2}
\equiv [M(N_{3},\kappa_2,3,k) \cdot M(N_{2},\kappa_1,\kappa_2,k) \cdot M(N_{1},3,\kappa_1,k)]_{2,2}
 \\
=1+\frac{1}{6}\left(\frac{k}{H} \right)^2 \times&\Biggl[
\frac{\kx_1-3}{\kx_1(\kx_1-2)}\left(-2(\kx_1-3)e^{N_1(\kx_1-2)-N_2\kx_1}
+3(\kx_1-2)e^{-2N_1}-\kx_1 e^{-2N_2} \right) \Biggr.
\\
&-\frac{2(\kappa_1-3)(\kappa_2-3)}{\kx_2(\kx_1-2)}
\left(e^{N_3\kx_2}-e^{N_2\kx_2} \right)
\left(e^{-2N_2-N_3\kx_2}-e^{N_1(\kx_1-2)-N_2\kx_1-N_3\kx_2} \right)
 \\
\Biggl.
&+\frac{\kx_2-3}{\kx_2(\kx_2-2)}\left(-2(\kx_2-3)e^{N_2(\kx_2-2)-N_3\kx_2}
+3(\kx_2-2)e^{-2N_2}-\kx_2 e^{-2N_3} \right)
\Biggr].
\end{split}
\end{equation}
For $\kx_2=3$ the second ``pulse" is eliminated and only the first 
term in the bracket survives, while for $\kx_1=3$ the first ``pulse" 
is eliminated and the last term
survives. For $\kx_1,\kx_2\not=3$ there is a
mixing term, which indicates that the effects of the various ``pulses" are not
simply additive, even within this approximation.

\begin{figure}[t!]
\centering
\includegraphics[width=0.4\textwidth]{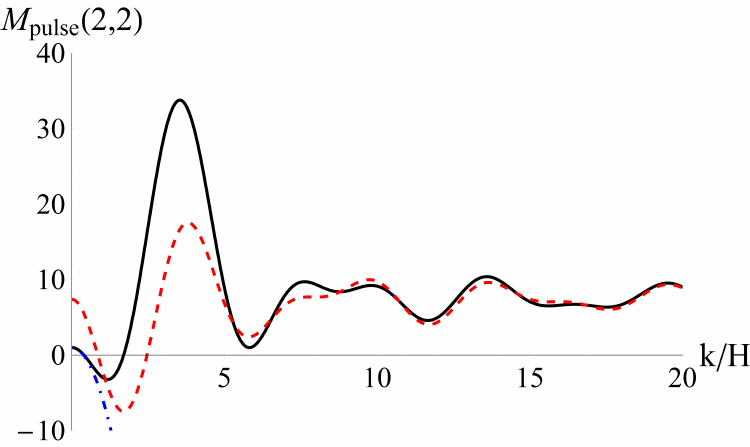}
\hspace{0.5cm}
 \includegraphics[width=0.4\textwidth]{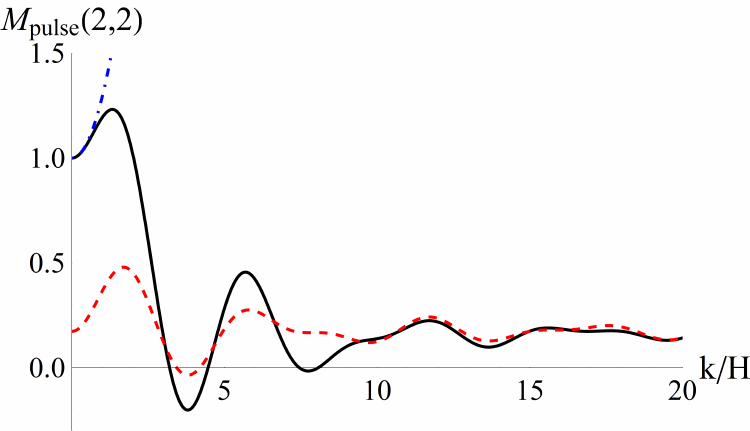} 
\caption{
The (2,2)-component of the matrix $M_{pulse}$ defined in eq. (\ref{Mpulse}) (solid curve),  
along with the approximations for large $k/H$ (dashed curve) and small $k/H$ (dot-dashed curve).
Left plot: $N_1=0$, $N_2=0.5$, $\kx=-5$. Right plot: $N_1=0$, $N_2=0.5$, $\kx=10$.
}
\label{Matrixpulse}
\end{figure}

The oscillatory behaviour of the solutions can be observed in the components of the
matrix $M_{pulse}$ defined in eq. (\ref{Mpulse}). In fig. \ref{Matrixpulse} we
depict the (2,2)-component of this matrix (solid lines) 
for  $N_1=0$, $N_2=0.5$. This component gives the leading contribution to the
power spectrum. The left 
plot corresponds to a negative-friction ``pulse" with $\kx=-5$ that causes the 
enhancement of the spectrum. The right plot is obtained for positive friction 
$\kx=10$ that leads to suppression.
The asymptotic expansions of this component for large $k/H$ (dashed curve), given by 
eq. (\ref{Mpulseinf}), and small $k/H$ (dot-dashed curve), given by eq. (\ref{Mpulse0}), 
are also plotted. Oscillatory behaviour is observed, associated with interference patterns
from two almost equal frequencies corresponding to $\delta k=e^{N_1}\pi$ and 
$\delta k=e^{N_2}\pi$. It is interesting that the oscillatory frequencies are correctly
reproduced by the asymptotic expansion even for small $k/H$.
Another feature that can be observed is the strong decrease of $M_{pulse}$
for $k/H \gta \kappa/2$ for positive $\kx$, in agreement with the discussion at the end
of the previous subsection.

\subsection{The integral of $f(N)$} \label{fNN}

The form of $f(N)$ that we assume in our discussion should result from the 
time evolution of the parameters $\exx_H$ and $\eta_H$. 
We saw at the end of subsection \ref{general} that the integral of this function 
over $N$ is constrained by eq. (\ref{integra}).
In this subsection we discuss the type of field evolution, as given by the function 
$\varphi_{,N}$, that is consistent with 
our approximate treatment.

The Mukhanov-Sasaki equation (\ref{eomv}) implies that the Wronskian of each Fourier 
mode of its solution 
\be
W[v_k] = -i(v_k\, v_k^{*\prime}-v_k^*v_k^{\prime})
\label{wronskian} \ee
remains constant during the evolution. 
Here a prime denotes a derivative with respect to conformal time $\tau=-e^{-N}/H$.
The solution of eq. (\ref{eomv}) plays the role of the mode function 
in the canonical quantization of the field $v$. For the Bunch-Davies 
vacuum, the initial condition at early times, when 
$k^2 \gg z''/z$, is chosen such that the mode function has the 
standard form in Minkowski spacetime. Selecting positive-energy 
solutions fixes the sign of the Wronskian to be positive, while the appropriate 
normalization results in $W[v_k]=1$.
This choice is automatically preserved at later times if $v_k$ is a solution of 
eq. (\ref{eomv}). This can be seen by multiplying eq. (\ref{eomv}) by $v_k^*$ and
subtracting the conjugate of the same equation multiplied by $v_k$.

We have based our analysis on the curvature perturbation $R_k$, related to $v_k$ through
$R_k=-v_k/z$, with $z=e^N \varphi_{,N}$. The consistency of 
our approximation of describing $f(N)$ through 
a sequence of ``pulses" implies a specific form of $\varphi_{,N}$ during the
evolution through the strong features in the potential. We can deduce this form
by considering the Wronskian of $R_k$
\be
W[R_k] = -i(R_k\, R_k^{*\prime}-R_k^*R_k^{\prime})
=\frac{W[v_k]}{z^2}=\frac{1}{z^2}.
\label{wronskianR} \ee
The solution (\ref{solBessel}) gives 
\be
W[R_k]\propto
i\left( C_p C_m^*-C_m C_p^* \right) \exp((1-\kx)N).
\label{WRprop} \ee
Consistency with eq. (\ref{wronskianR}) requires that $\varphi_{,N}\propto \exp((\kx-3)N/2)$. The inflaton ``velocity" must grow fast with $N$ 
for $\kx>3$, and decay for $\kx <3$. This is the behaviour observed 
in fig. \ref{friction}.
We have already mentioned that any function $f(N)$
can be reconstructed as a sequence of very short ``pulses" of variable height $\kx$.
For small $N$ the change of 
$\varphi_{,N}$, starting from some initial value at $N=0$, is linear in $N$ with a slope proportional to $\kx$.
Thus, by breaking $f(N)$ into many ``pulses"
one can obtain the required evolution of $\varphi_{,N}$
as a function of $N$.
In this sense our analysis is very general. 
For consistency, of course, 
the deduced evolution must result from an appropriate inflaton potential.

Our main aim is to obtain an intuitive understanding
of the form of the spectrum by focusing on the gross properties of $f(N)$.
Let us consider a feature in the evolution resulting from two successive 
``pulses" with heights $\kx_1$ and $\kx_2$, between early and late slow-roll
regimes with $\kx=3$. The solution after the feature
is traversed is given by eq. (\ref{solBessel3}) with 
\be
\begin{pmatrix}  
		 C_{p}\\
		  C_{m}
\end{pmatrix}=
M(N_3,\kappa_2,3,k)
\cdot M(N_2,\kappa_1,\kappa_2,k)
\cdot M(N_1,3,\kappa_1,k)
\begin{pmatrix}	  
		 1\\
		  i
\end{pmatrix},
\label{spectrumtwopulse}	  
\ee
where the matrix $M$ is given by eq. (\ref{Mab}). 
Before the ``pulse" we have $i(C_p C_m^*-C_m C_p^*)/2=1$, 
while after the ``pulse" one finds 
\be
\frac{i}{2}(C_p C_m^*-C_m C_p^*)= e^{-(n_2-n_1)(\kx_1-3)-(n_3-n_2)(\kx_2-3)}.
\label{wronskiantwopulse} \ee
The exponent is exactly (minus) the integral of $f(N)-3$.
By comparing the Wronskian $W[R_k]$ at late and early times (before and after 
the ``pulse"), it becomes
clear the the quantity (\ref{wronskiantwopulse}) is equal to the ratio 
$\left(\varphi^2_{,N}\right)_e/\left(\varphi^2_{,N}\right)_l$, with both
quantities being constant. 
In this way we reproduce the result of eq. (\ref{integra}), under our assumption
that $H_l/H_e \simeq 1$.

Let us summarize the basic points:
According to our assumptions, the system is in a slow-roll regime during an
early and a late period,
with values of the Hubble parameter that we have approximated as equal. 
We can 
assume that the values of $\varphi_{,N}$ are also approximately equal during
these periods. These 
assumptions isolate the effect of the strong feature in the 
intermediate part of the evolution from the 
properties in the slow-roll regimes.
During the intermediate part the inflaton ``velocity" $\varphi_{,N}$
changes fast, 
by growing or decaying depending on the sign of $f(N)-3$, as observed 
in fig. \ref{friction}.
The integral of $f(N)-3$ over $N$ must vanish for $\varphi_{,N}$ to
have equal values at early and late times.  
For realistic situations one must take into account the breaking of 
scale invariance in the slow-roll regimes as well. 
However, these are included in the standard slow-roll
analysis and are not of interest to us here.

Finally, it can be checked through the asymptotic form of the Bessel functions 
that for both $k\to 0$ and $k \to \infty$, and for a vanishing integral of
$f(N)-3$, we have $(C_p,C_m)=(1,i)$ at all times during the evolution.
This indicates that the low- and high-$k$ modes are not affected by the presence of the feature. As a result the scale-invariant form of the 
spectrum is modified only for a finite range of wavenumbers $k$.

\subsection{The form of the spectrum} \label{form}

In this subsection we consider three examples of spectra that display the features 
 we discussed in the previous subsection. The range of possible spectra is large, as 
 we do not focus on a particular underlying model, but simply consider various forms
 of the function $f(N)$ defined in eq. (\ref{AN}). 
We assume that the integral of $f(N)-3$ over $N$ vanishes, so that the 
spectrum is scale invariant with the same amplitude 
for very low and very high wavenumbers $k$.
We focus only on the relative enhancement of the spectrum at intermediate scales
as a result of the presence of strong features in the underlying inflaton evolution. As
the absolute scale of the spectrum is not of interest for our discussion, we set 
$A=1$ in eq. (\ref{solBessel3}). We discuss next three particular examples of the form of the friction function $f(N)$.

\begin{figure}[t!]
\centering
\includegraphics[width=0.9\textwidth]{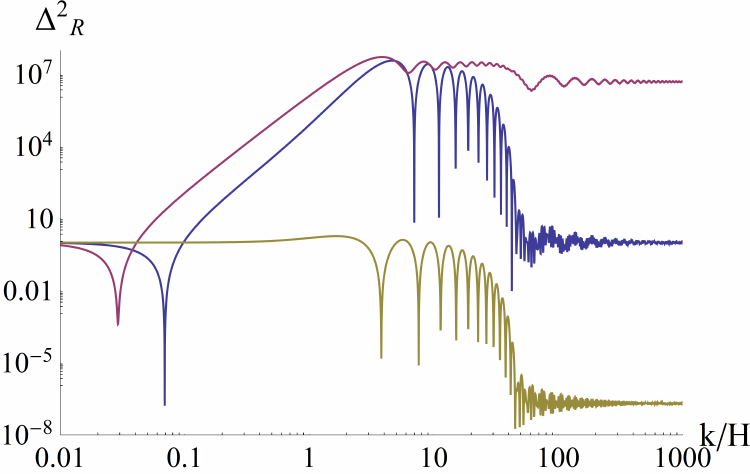}
\\
\includegraphics[width=0.5\textwidth]{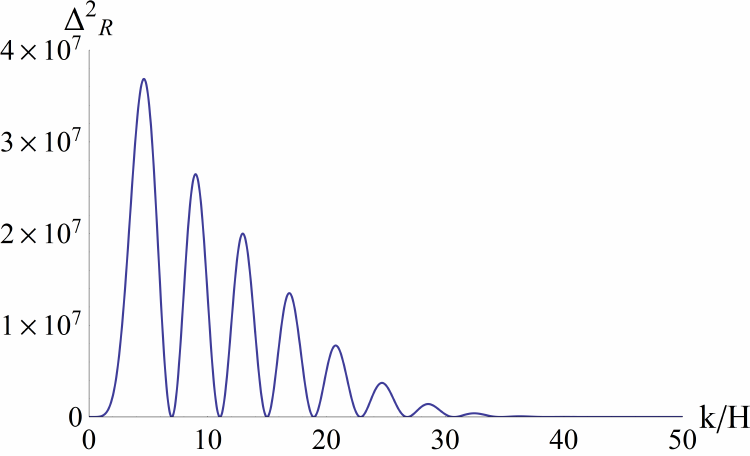}
\hspace{0.5cm} 
 \includegraphics[width=0.4\textwidth]{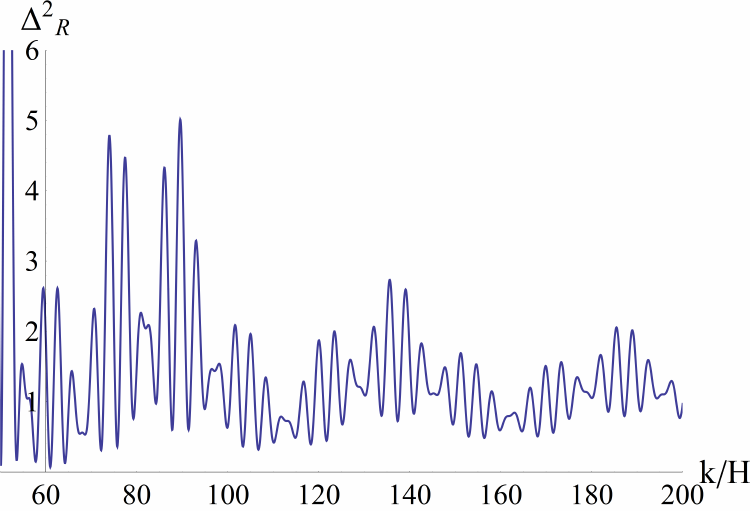} 
\caption{Middle curve: Spectrum resulting from a double ``pulse" with 
$N_1=0$, $N_2=0.2$, $N_3=2.77$, $\kx_1=80$, $\kx_2=-3$ (Ex. 1). Upper curve: Spectrum
resulting from a negative-friction single ``pulse" with $\kx_2=-3$
between $N_2=0.2$, $N_3=2.77$. Lower curve: Spectrum resulting from a positive-friction 
single ``pulse" with $\kx_1=80$ between $N_1=0$, $N_2=0.2$.
}
\label{spectrum1}
\end{figure}

In our first example (Ex. 1) the spectrum 
results from a function $f(N)$ of the qualitative form depicted in the first line
of fig. \ref{friction} and displayed explicitly in fig. \ref{cartoon}. The
feature consists of a positive-friction ``pulse" with $\kx_1=80$ in the 
interval between $N_1=0$ and $N_2=0.2$, followed by a negative-friction ``pulse" with
$\kx_2=-3$ in the interval between $N_2=0.2$ and $N_3=2.77$. The value of the spectrum
for a given value of $k/H$ is equal to $|C_m|^2$, where
$(C_p, C_m)$ is given by eq. (\ref{spectrumtwopulse}).
The result is depicted by the middle curve
of the top plot in fig. \ref{spectrum1},
 in the $k$-range $k/H=10^{-2}-10^{3}$ that corresponds to $N\simeq -4.6$ up to  $6.9$. 
In the same figure we also display the spectra that would result from a 
single ``pulse". These are computed from the expression 
\be
\begin{pmatrix}  
		 C_{p}\\
		  C_{m}
\end{pmatrix}=
 M(N_2,\kappa_1,3,k)
\cdot M(N_1,3,\kappa_1,k)
\begin{pmatrix}	  
		 1\\
		  i
\end{pmatrix},
\label{spectrumonepulsep}	  
\ee
for the positive-friction ``pulse" (lower curve in fig. \ref{spectrum1}),
and 
\be
\begin{pmatrix}  
		 C_{p}\\
		  C_{m}
\end{pmatrix}=
M(N_3,\kappa_2,3,k)
\cdot M(N_2,3,\kappa_2,k)
\begin{pmatrix}	  
		 1\\
		  i
\end{pmatrix},
\label{spectrumonepulsen}	  
\ee
for the negative-friction ``pulse" (upper curve in fig. \ref{spectrum1}).
As we discussed in the previous subsection, the fact that the integral 
of the function $f(N)-3$ over $N$ does not vanish for these cases means that
the quantity $\varphi_{,N}^2$ changes across the ``pulse" 
by a factor equal to the exponential of this integral. The two slow-roll
regimes are quite distinct in this case and the effect of the ``pulse" is
not clear.
 We display the spectra because 
they provide intuition on the features appearing in the 
two-``pulse" 
spectrum, for which the integral of $f(N)-3$ vanishes. Details for the latter
are presented in the next two plots of 
fig. \ref{spectrum1}, for two successive $k/H$ ranges 
on a linear horizonal axis. Notice the huge difference in the scale 
of the vertical axis in the two plots.

\begin{figure}[t!]
\centering
\includegraphics[width=0.9\textwidth]{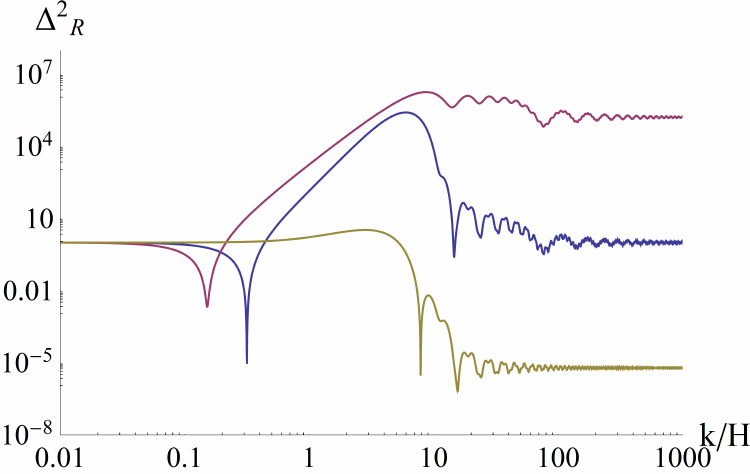}
\\
\includegraphics[width=0.5\textwidth]{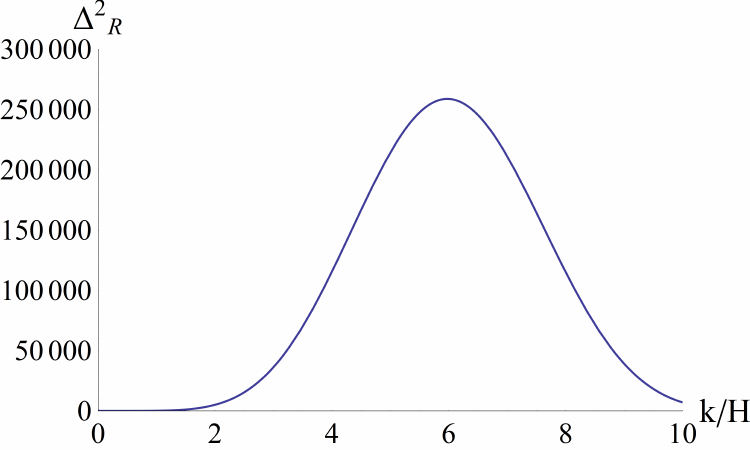}
\hspace{0.5cm}
 \includegraphics[width=0.4\textwidth]{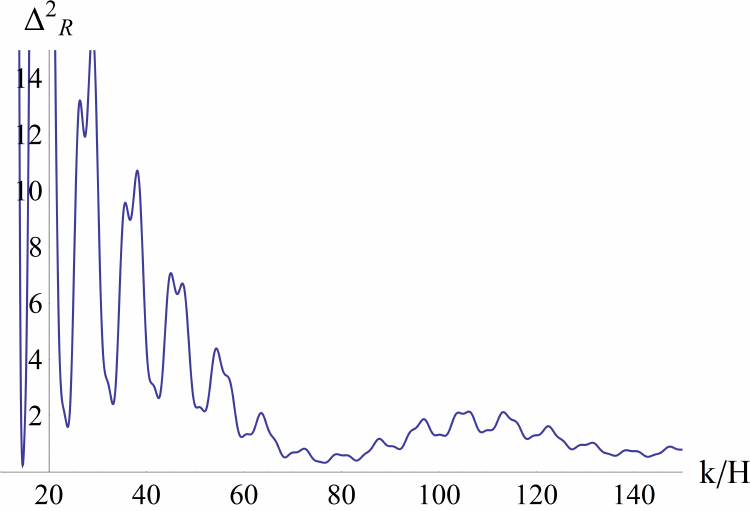} 
\caption{
Middle curve: Spectrum resulting from a double ``pulse" with 
$N_1=0$, $N_2=1$, $N_3=3$, $\kx_1=15$, $\kx_2=-3$ (Ex. 2). Upper curve: Spectrum
resulting from a negative-friction single ``pulse" with $\kx_2=-3$
between $N_2=1$, $N_3=3$. Lower curve: 
Spectrum resulting from a positive-friction 
single ``pulse" with $\kx_1=15$ between $N_1=0$, $N_2=1$.
}
\label{spectrum2}
\end{figure}

\begin{figure}[t!]
\centering
\includegraphics[width=0.9\textwidth]{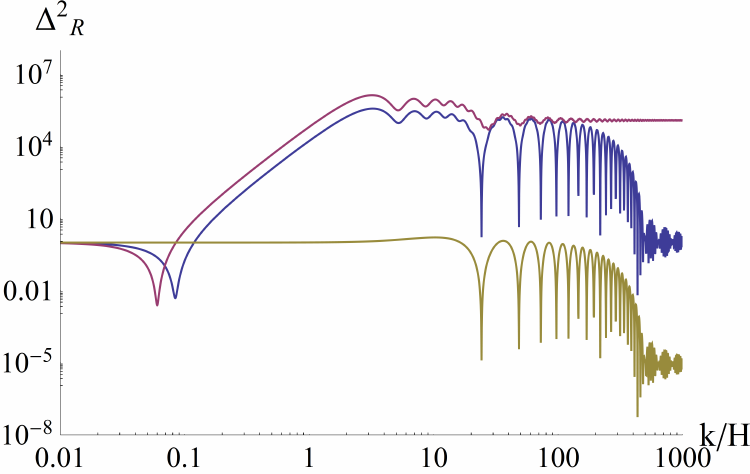}
\\
\includegraphics[width=0.5\textwidth]{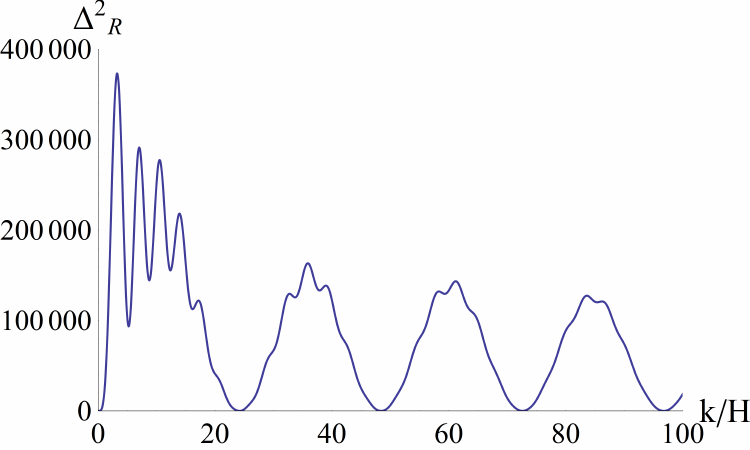}
\hspace{0.5cm}
 \includegraphics[width=0.4\textwidth]{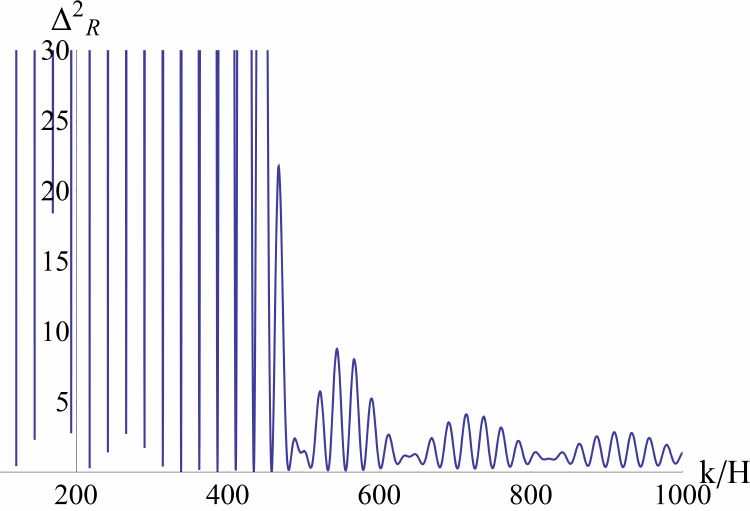} 
\caption{
Middle curve: Spectrum resulting from a double ``pulse" with 
$N_1=0$, $N_2=1.95$, $N_3=2.05$, $\kx_1=-3$, $\kx_2=120$ (Ex. 3). Upper curve: Spectrum
resulting from a negative-friction single ``pulse" with $N_1=0$, $N_2=1.95$, $\kx_1=-3$. Lower curve: Spectrum resulting from a positive-friction 
single ``pulse" with $\kx_2=120$ between $N_2=1.95$, $N_3=2.05$.
}
\label{spectrum3}
\end{figure}

Several features of the spectra are apparent in these plots:
\begin{enumerate}
\item
The two-``pulse" spectrum has a first minimum at a value of $k/H$ well 
approximated by the positive root of the polynomial of eq. (\ref{Mpulse0}). 
\item
The subsequent strong increase of the spectrum results from the effect of the 
negative-friction ``pulse". The spectrum reaches a maximal value comparable
to that of the negative-friction single-``pulse" spectrum.
A rough estimate can be obtained from the asymptotic value of the  
single-``pulse" spectrum, which is $\exp((N_3-N_2)(3-\kx_2))={\cal O}(10^7)$.
\item
The envelope of the 
positive-friction single-``pulse" spectrum (lower curve) displays a sharp drop to
almost zero at a characteristic value of $k/H$. 
As we discussed earlier, we expect that the positive friction will affect most strongly
the high-$k$ modes. A more quantitative estimate can be made by observing that the 
matrix $M$ of eqs. (\ref{Mab}) involves the Bessel functions $J_{\pm \kx_1/2}$ and
$J_{\mp 1 \pm \kx_1/2}$. For large $\kx_1$ these functions have a zero at 
a value of their argument roughly equal to $\kx_1/2$. The relevant argument in our
case is $e^{-\bar{N}}k/H$, with $\bar{N}\simeq(N_1+N_2)/2$. 
Thus, we expect the spectrum to approach zero at 
$k/H\simeq e^{\bar{N}} \kx_1/2 \simeq 44$, consistently with what is observed.
\item
For $k/H\to \infty$, all three spectra become asymptotically constant, with 
values given by the 
exponential of the integral of $f(N)-3$ over $N$. For the middle spectrum, we
have fine-tuned this integral to zero, so that the spectrum returns to the 
value 1 to which we have normalized the spectrum for $k\to 0$.
\item
Apart from the main features that we described above, which are consistent with
the general expectations \cite{ozsoy}, the spectra display 
oscillations with characteristic scales. As we discussed in the previous 
subsection, the asymptotic expansions of 
eqs. (\ref{Mpulseasympt}) indicate that the spectrum should oscillate with periods
$\delta k/H\simeq e^{N_1}\pi= 3.1$,
$\delta k/H\simeq e^{N_2}\pi= 3.8$ and $\delta k/H\simeq e^{N_3}\pi=50$. 
These characteristic modes, as well as interference patterns between them, are 
visible in the bottom plots of fig. \ref{spectrum1}. 
\end{enumerate}
The most important conclusion that can be drawn from this example is that
strong features in the background evolution can induce a 
spectrum of fluctuations which displays, apart from an enhancement by
several orders of magnitude, strong oscillatory patterns. This is clearly
visible in the bottom left plot of fig. \ref{spectrum1}.

We turn next to our second example (Ex. 2). 
The oscillatory features in the spectrum are less pronounced for
different forms of the ``pulses". Reducing the height of the positive-friction ``pulse"
leads to a suppression of the spectrum at smaller values of $k/H$. As a result the 
oscillatory patterns may be confined within the high-$k$ part of the spectrum, which does
not get enhanced.
This is visible in fig. \ref{spectrum2}, where we plot the spectrum 
in the $k$-range, $k/H=10^{-2}-10^{3}$ or from  $N\simeq -4.6$ up to  $6.9$.
The spectrum results from
a positive-friction ``pulse" with $\kx_1=15$ in the 
interval between $N_1=0$ and $N_2=1$, followed by a negative-friction ``pulse" with
$\kx_2=-3$ in the interval between $N_2=1$ and $N_3=3$.
The drop of the spectrum arising from only the
positive-friction ``pulse" is expected to appear at 
 $k/H\simeq e^{\bar{N}} \kx_1/2 \simeq 12$, where $\bar{N}\simeq (N_1+N_2)/2$.
Indeed, the small-$k$ region displays a large enhancement,
but the oscillations appear only at large values of $k/H$, at which the spectrum is
suppressed.
The bottom left plot of fig. \ref{spectrum2} shows that the enhanced
part of the spectrum is smooth in this case. 
The expected oscillatory modes with periods
$\delta k/H\simeq e^{N_1}\pi= 3.1$,
$\delta k/H\simeq e^{N_2}\pi= 8.5$ and $\delta k/H\simeq e^{N_3}\pi=63$, 
as well as interference patterns between them, are 
visible in the bottom right plot of fig. \ref{spectrum2}. 

Our third example (Ex. 3) demonstrates that 
spectra with a different structure can result from different forms of the function 
$f(N)$. More specifically, the positive- and negative-friction ``pulses" may occur in the 
reverse order compared to the one we assumed up till now. This is possible if the inflaton 
encounters a region of the potential with almost vanishing slope, as displayed
in the second line of plots in fig. \ref{friction}. The reduction of the field
``velocity" results in a period of positive values for the parameter $\eta$.
When the inflaton moves beyond this region its ``velocity" grows again, with $\eta$ taking negative values.
In fig. \ref{spectrum3} we plot the resulting spectra in the range  $k/H=10^{-2}-10^{3}$ ($N\simeq -4.6$ up to  $6.9$) considering an effective friction function $f(N)$ composed of
 a negative-friction ``pulse" with $\kx_1=-3$ in the 
interval between $N_1=0$ and $N_2=1.95$, followed by a strong positive-friction ``pulse" with
$\kx_2=120$ in the interval between $N_2=1.95$ and $N_3=2.05$.
The reduction of the spectrum is expected at a scale 
 $k/H\simeq e^{\bar{N}} \kx_2/2 \simeq 440$, where $\bar{N}\simeq (N_2+N_3)/2$.
The oscillatory patterns have characteristic periods
$\delta k/H\simeq e^{N_1}\pi= 3.1$,
$\delta k/H\simeq e^{N_2}\pi= 22.1$ and $\delta k/H\simeq e^{N_3}\pi=24.4$.
All these features, as well as strong interference patterns arising from the 
proximity of two characteristic periods, are visible in fig. \ref{spectrum3}.


\subsection{Analytical expressions for general $f(N)$} \label{analytfN}

In this subsection we derive analytical expressions for the curvature spectrum resulting from 
an arbitrary friction function $f(N)$.
We start by rewriting eq. (\ref{RN}) as 
\be
R_{k,NN}+3R_{k,N}+\frac{k^2}{e^{2N} H^2} R_{k}=
(3-f(N))R_{k,N}.
\label{ANg} \ee
We would like to compute the Green's function $G(N)$ for the operator in the lhs.
This function satisfies the equation 
\be
G_{k,NN}(N,n)+3G_{k,N}(N,n)+\frac{k^2}{e^{2N} H^2} G_{k}(N,n)=\delta(N-n).
\label{ANGr} \ee
The solution of eq. (\ref{ANg}) is
\be
R_k(N)=\Rb_k(N;1,i,3)+\int_{-\infty}^{\infty}G_k(N,n)\,(3-f(n))\,R_{k,n}(n)\, dn,
\label{solg} \ee
with 
\be
\Rb_k(N;1,i,3)=-\sqrt{\frac{2}{\pi}}\left( \frac{H}{k} \right)^{3/2}
 \left(i+  e^{-N}\frac{k}{H} \right) \exp \left(i e^{-N}\frac{k}{H}  \right)
\label{Rk0} \ee
the solution of the homogeneous equation, corresponding to $f(N)=3$.

The evolution is classical, so we must use the retarded Green's function, which 
satisfies $G_{k>}(N,n)=0$ for $n>N$. For $n<N$ the Green's function is 
\be
G_{k<}(N,n)= e^{-\frac{3}{2}\, N} \left(A(n) J_{3/2}\left( e^{- N}\frac{k}{H} \right) + B(n)\, J_{-3/2}\left( e^{-N}\frac{k}{H} \right)  \right).
\label{Grless} \ee
The total Green's function is continuous at $N=n$. Its first derivative has a discontinuity, obtained
by integrating eq. (\ref{ANGr}) around $N=n$. This gives $\partial G_{k<}(N,n)/\partial N|_{N=n}=1.$
Imposing these constraints results in
\begin{eqnarray}
A(n)&=&-\sqrt{\frac{\pi}{2}}e^{3n}\left(\frac{k}{H} \right)^{-3/2}
\left(\cos\left(e^{-n}\frac{k}{H} \right) +e^{-n}\frac{k}{H}\sin\left(e^{-n}\frac{k}{H} \right)\right)
\label{An} \\
B(n)&=&\sqrt{\frac{\pi}{2}}e^{3n}\left(\frac{k}{H} \right)^{-3/2}
\left(e^{-n}\frac{k}{H}\cos\left(e^{-n}\frac{k}{H} \right) -\sin\left(e^{-n}\frac{k}{H} \right)\right).
\label{Bn} \end{eqnarray}

Despite its simple form, 
it is difficult to find solutions of eq. (\ref{solg}). However, the equation becomes
simpler for $N\to \infty$, which is the limit of interest for the late-time spectrum.
From eq. (\ref{Grless}) we obtain 
\be
G_{k<}(N,n)\to -\sqrt{\frac{2}{\pi}}\left( \frac{H}{k} \right)^{3/2}\,B(n)
\label{Grlesslim} \ee
in this limit.
Eq. (\ref{solg}) now becomes
\be
R_k(\infty)=\Rb_k(\infty;1,i,3)
 -\sqrt{\frac{2}{\pi}}\left( \frac{H}{k} \right)^{3/2}
\int_{-\infty}^{\infty}(3-f(n))\,B(n)\,R_{k,n}(n)\, dn,
\label{solg2exact} \ee
with
\be
\Rb_k(\infty;1,i,3)= -i \sqrt{\frac{2}{\pi}}\left( \frac{H}{k} \right)^{3/2}.
\label{inf} \ee
Even though an analytical solution of this equation is not available, some conclusions about its
form can be drawn when the function $f(N)$ displays strong features. The clearest example 
is a feature that can be approximated through a $\delta$-function centered at $N_1$.
The integration over $n$ results in an expression that includes $\sin(e^{-N_1}k/H)$ and 
$\cos(e^{-N_1}k/H)$, producing  oscillatory patterns. A similar conclusion can be reached if $f(N)$ involves sharp step-like features
approximated through $\Theta$-functions, as we discussed in the previous
subsection. These patterns are expected to become less 
prominent when the features in $f(N)$ become smoother.

\begin{figure}[t!]
\centering
\includegraphics[width=0.4\textwidth]{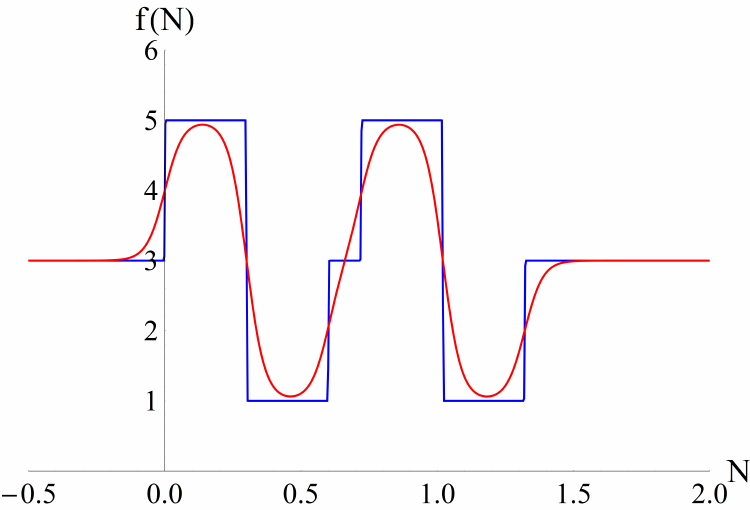}
\hspace{0.5cm}
 \includegraphics[width=0.5\textwidth]{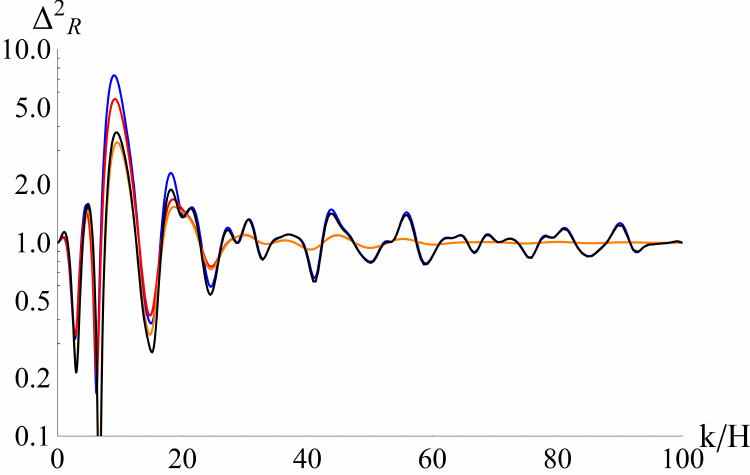} 
\\
\vspace{0.5cm}
\includegraphics[width=0.4\textwidth]{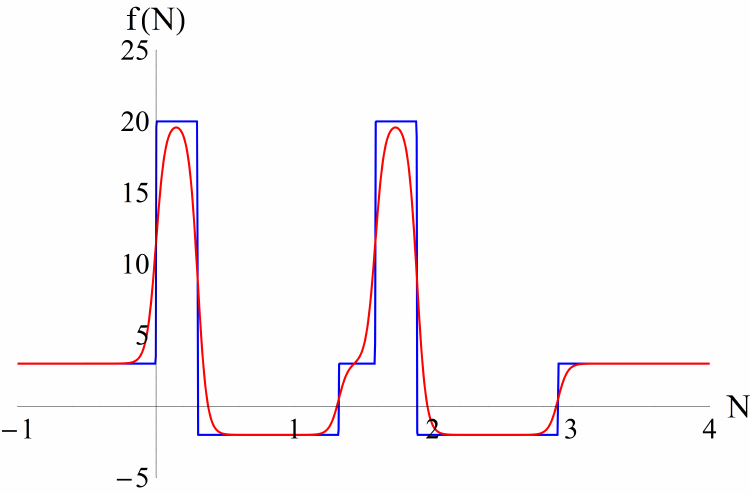}
\hspace{0.5cm}
 \includegraphics[width=0.5\textwidth]{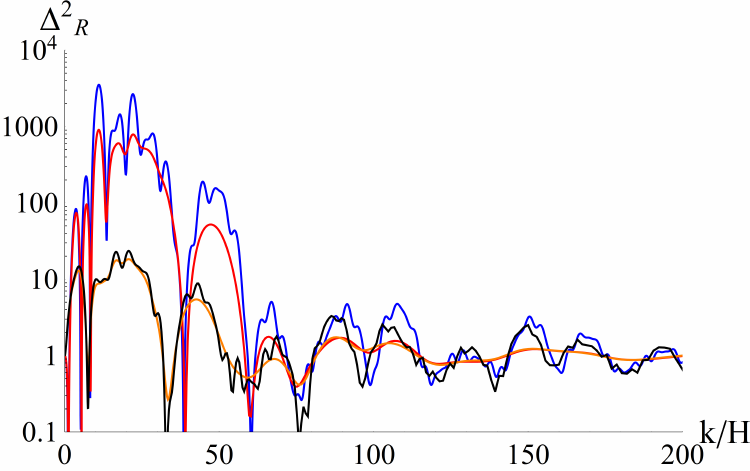} 
\caption{
Curvature power spectra for various forms of the friction function $f(N)$ (blue and red curves),
compared to predictions by
eq. (\ref{solg2}) (black and orange curves). 
}
\label{extragreen}
\end{figure}

An approximate expression, which can be considered as the first step in an iterative solution of
the above equation, can be obtained if we replace the full solution $R_k(n)$ in the integral
with the solution for $f(n)=3$, given by $\Rb_k(n,1,i,3)$. We have 
\be
\Rb_{k,n}(n,1,i,3)=\sqrt{\frac{2}{\pi}}\left( \frac{k}{H} \right)^{1/2}e^{-2n}
\left( i \cos\left( e^{-n}\frac{k}{H} \right)-\sin\left( e^{-n}\frac{k}{H} \right) \right).
\label{apprderiv} \ee  
Combining the above expressions, we obtain
\begin{eqnarray}
&&R_k(\infty)=\Rb_k(\infty;1,i,3) \times
\nonumber \\
&&\,\left\{ 1-i \frac{H}{k}
\int_{-\infty}^{\infty}(3-f(n))\,e^n 
\left[ e^{-n}\frac{k}{H}\cos\left(e^{-n}\frac{k}{H} \right) -\sin\left(e^{-n}\frac{k}{H} \right)\right]
\,
\left[ i \cos\left( e^{-n}\frac{k}{H} \right)-\sin\left( e^{-n}\frac{k}{H} \right) \right]
\, dn \right\},
\nonumber \\
&~&
\label{solg2} \end{eqnarray}
This result is expected to be valid only for cases without a large enhancement of the 
spectum. However, it is a compact expression that can be used in order to deduce 
the expected oscillatory patterns for a general form of $f(N)$.

In fig. \ref{extragreen} we examine the validity of eq. (\ref{solg2}) for $f(N)$ with 
sharp and smooth features.
In the left plot of the first line we depict the sharp and smoothed version 
(blue and red lines, respectively)
of a friction function with moderate deviations from 3. In the right plot we depict
the corresponding exact spectra (blue and red lines, respectively), as well as the ones computed through
the approximate expression of eq. (\ref{solg2}) (black and orange lines, respectively).
It is apparent that this expression captures with very good accuracy 
the complicated oscillatory patterns: the blue and black lines, as well as the red and orange lines,
are in very good agreement. Small deviations from the exact solutions
are observed when the amplitude becomes large.
In the second line we repeat the calculation for a function $f(N)$ with very strong features
that result in a large enhancement of the spectrum. The limitations of eq. (\ref{solg2})
in capturing the magnitude of the enhancement become apparent. The agreement between 
black-blue lines and red-orange lines is good only for large values of $k/H$, for which 
the deviations from the scale-invariant spectrum are small (but still of {\cal O}(1)). 
Clearly, higher orders in the iterative solution of eq. (\ref{solg2exact}) are needed in order
to capture the strong enhancement of the spectrum at small values of $k/H$.

Despite the limited range of validity of eq. (\ref{solg2}), it is interesting that the predicted
oscillatory pattern appears in good agreement with the exact result in all cases. This 
indicates that the characteristic frequencies are determined by the convolution of the
friction function $f(n)$ with the functions $\sin(e^{-n}k/H)$ and 
$\cos(e^{-n}k/H)$ in eq. (\ref{solg2exact}), even if $R_k(n)$ deviates strongly from 
the unperturbed solution $\Rb_k(n;1,i,3)$ of eq. (\ref{Rk0}). In this respect, eq. (\ref{solg2})
provides the
means for estimating the frequencies that appear in the power spectra for a general form
of the friction function $f(N)$.

We also note that, 
as we saw in the previous subsection, the maximal enhancement of the spectrum for a single ``pulse" 
can be estimated through 
the exponential of the integral of $3-f(N)$ over the range that this function is positive.
For patterns involving several ``pulses", the enhancement depends on their relative position 
\cite{Kefala:2020xsx}. However, the above estimate can be used as a (rough) guide for 
the maximal enhancement of the spectrum for the optimal position of the ``pulses".

\section{Primordial black holes and induced  gravitational waves}
 \label{blackholesGW}

\subsection{Specific inflationary models} \label{Secinfmodels}

Based on the discussion of section \ref{stepalpha} 
that motivated the use of the framework of $\alpha$-attractors, 
we assume the following form for the function $F$:
\begin{equation}
F(x)= F_0 \left(x+ \sum_{i=1}^n c_i\tanh\left(d(x-x_i)\right) \right) \, .
\end{equation}
The corresponding inflationary potential for the field $\varphi$ in the Einstein frame
\begin{equation} \label{Vall}
V(\varphi)=F^2\left(\tanh\frac{\varphi}{\sqrt{6}}\right)\
\end{equation}
 features $n$ step-like transitions.
 (All dimensionful quantities are given in units of $\mpl$.)
Such a potential can lead to an enhancement of the power spectrum of scalar perturbations \cite{Kefala:2020xsx} at particular scales, which can be sufficiently large 
to trigger PBH formation and induce detectable GWs. 
In addition, the shape of the scalar power spectrum around its peak is characterized 
by an oscillatory pattern that can be inherited  by the tensor power spectrum.  
We will discuss these notable phenomenological implications of potentials with steps 
in the next subsections.

\begin{figure} 
   \centering
   \includegraphics[width=.49 \linewidth]{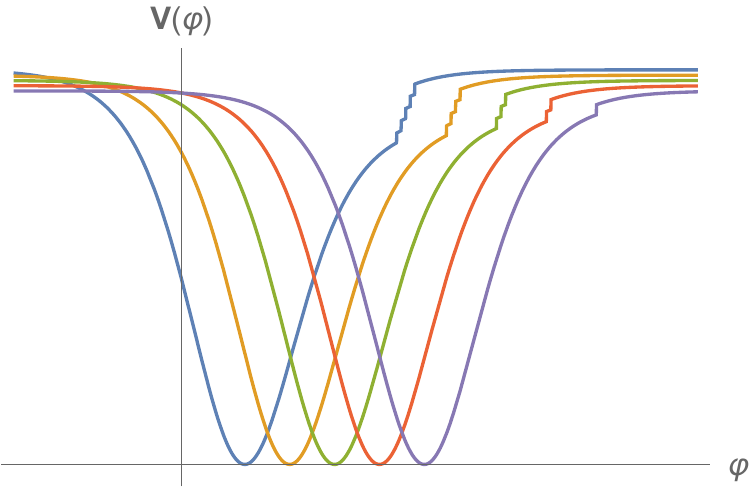}
    \includegraphics[width=.49 \linewidth]{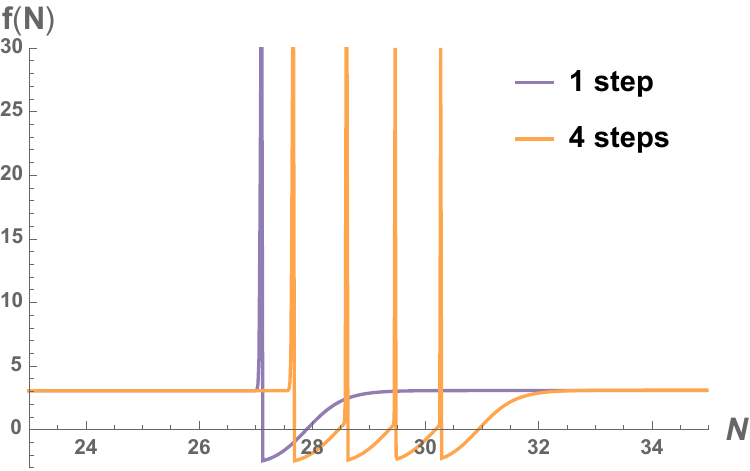}
  \caption{\label{FigVstepsCollection}~~  
Left panel: The inflationary potentials described by eq. (\ref{Vall}), arbitrary placed on the $\varphi$ axis in order 
to make the step-like structure visible. 
Right panel: The function $f(N)$ of eq. (\ref{AN})
in terms of the number of efoldings for inflationary  
potentials with one and four steps.
   }
 \end{figure}

The enhancement induced by a step has an upper bound corresponding roughly to a multiplicative factor 
$\exp({-\Delta N(\kappa-3)})$, see eq.(\ref{enhancement}). Here $\Delta N$ is the
interval during which the value
$\kappa$ of the effective-friction term 
(\ref{AN}) is smaller than the value $\kappa= 3$ that results 
in a scale-invariant spectrum. Negative values of $\kappa$ are realized 
when the background inflaton ``decelerates" on the lower plateau, after a sharp transition
through a step in the potential. During this stage, which lasts a few efoldings, we have
$\varphi_{,NN}\simeq -3 \varphi_{,N}$ and $\kappa\simeq -3$. 
As a result, 
a single step generally enhances the scalar power spectrum by roughly two or three orders of
magnitude. 
However, it is possible that the potential includes several step-like features.
 In fig. {\ref{FigVstepsCollection}} we plot a set of specific examples of inflationary potentials with steps, described by eq. (\ref{Vall}).
  These potentials yield 50 to 60 efoldings after the crossing of the CMB scale ($k=0.05\, \text{Mpc}^{-1}$)
  and a spectral index value $n_s=0.969$, 
  within the $68 \%$ CL range of Planck \cite{Akrami:2018odb}.
The parameters of these models are $c_i=7\times 10^{-3}$.
We consider from one ($n=1$) up to five steps ($n=5$)  in eq. (\ref{Vall}), placed at $\varphi_1=5.7$, $\varphi_2=5.55$, $\varphi_3=5.4$, $\varphi_4=5.25$, $\varphi_5=5.1$, respectively. The value of $F_0$ is adjusted each time 
in order to be consistent with the measured amplitude of the spectrum at the CMB scale.
For the initial value of the inflaton field we choose $\varphi_\text{CMB}=6.33$, 
so as to obtain appropriate values for  the spectral index $n_s$ and the 
number of efoldings $N$. 
The choice of the value of the parameter $d$ is not crucial, as long as it is taken 
sufficiently large for the transition through
the steps to occur quickly, but continuously. Typical values are of order $10^3-10^{5}$.

\begin{figure} 
   \centering
   \includegraphics[width=.49 \linewidth]{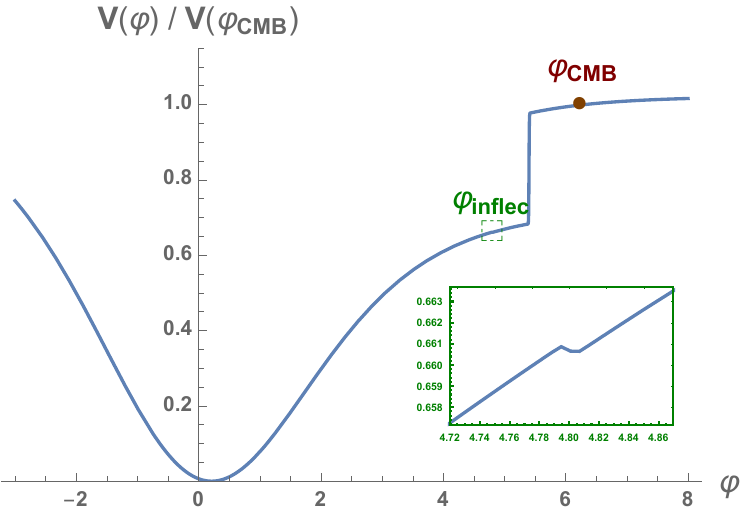}
    \includegraphics[width=.49 \linewidth]{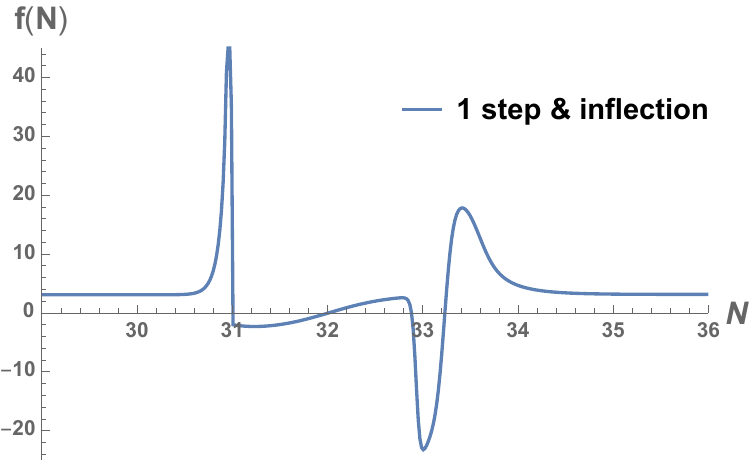}
  \caption{\label{FigVstepInflection}~~  
Left panel:  The inflationary potential with one step and an inflection point
as a function of $\varphi$  in Planck units. 
The model gives $n_s=0.968$ for an 
initial field value $\phi_\text{CMB}=6.17\, M_\text{Pl}$. 
The inflection point at $\phi_\text{inflec}=4.8\, M_\text{Pl}$ is clearly visible
through the magnification of the potential in the box.  
Right panel: The effective-friction function $f(N)$ during the
part of the evolution in which it deviates from the standard value 
$f(N)\simeq 3$.
   }
 \end{figure}

We also examine the inflationary dynamics of models that feature both a step and a near-inflection point. 
The production of PBHs and induced GWs due to the presence of a near-inflection point in the framework of $\alpha$-attractors has been studied in \cite{Dalianis:2018frf, Dalianis:2019asr, Dalianis:2020cla}.
Such models result in  a significant enhancement of the scalar power spectrum, while 
the presence of a step-like feature  adds a prominent oscillatory pattern around the peak value. 
In fig. \ref{FigVstepInflection} we plot an example of such a potential, 
within the $\alpha$-attractor framework, with parameters $c_1=8.70\times 10^{-2}$, $c_2=-2.77\times10^{-4}$. The step is placed at  $\varphi_1=5.4$ and a shallow nearly-inflection point exists at $\varphi_2=4.8$. 
The spectral index value for this model is $n_s=0.968$, within the $68 \%$ CL region of Planck \cite{Akrami:2018odb}. 
The number of efoldings after the crossing of the CMB scale is $N=51$ 
for an initial field value $\varphi_\text{CMB}=6.17$. 
In figs. \ref{FigVstepsCollection} and \ref{FigVstepInflection} we also plot 
the function $f(N)$ that determines key characteristics of the scalar power spectrum, 
such as the amplitude and the oscillatory pattern, as discussed in section \ref{notes}.

In the following subsections we examine the cosmological implications for 
PBH formation and GW production arising from the 
amplification of the scalar spectrum by the step-like features in 
the potential (\ref{Vall}). 
Remarkably, models of this type yield striking predictions for 
the induced GWs that render them testable by the forthcoming GW detection experiments.

\subsection{Primordial black holes}

Inflationary potentials with steps enhance the amplitude of the primordial density perturbations at particular scales and might lead to gravitational collapse and PBH production. We review briefly observational bounds on the PBH abundance, relevant for our analysis.

  In the largest part of the mass spectrum there are stringent upper bounds on $\Omega_\text{PBH}/\Omega_\text{DM}$ 
  arising from observational constraints, see fig.~\ref{figfPBH} for  monochromatic PBH spectra.
Light PBHs are constrained by the extra-galactic gamma ray background (EGB); 
black holes of mass above $10^{17}$g are subject to constraints from 
gravitational lensing of stars 
by Subaru (HSC), Ogle (O), EROS (E) and MACHO (M), microlensing of supernova (SN) and other experiments.
The CMB anisotropies measured by Planck (PA) constrain the PBHs with masses above $10^{33}$g. 
In the large-mass region there are also constraints from accretion limits in X-ray and radio observations 
 and X-ray binaries (XB),  
 and dynamical limits from disruption of wide binaries (WB)
 and survival of star clusters in Eridanus II (Er). 
Advanced LIGO/Virgo searches for compact binary systems with component masses 
in the range $0.2-1 M_\odot$ find no GW events.
For a detailed discussion and references on the PBH constraints we 
refer the reader to \cite{Carr:2020gox}.

The maximal value of the PBH abundance can be achieved 
in the mass range $M_\text{PBH}\sim 10^{-15}-10^{-10} M_\odot$.
In this work we focus on this mass window that can be tested by near-future GW experiments, such as LISA.
Nonetheless, the parameters of the same inflationary model with 
step-like features can be adjusted in order to generate PBHs in other mass 
windows, such as the  ${\cal O}(10)$ solar mass window that is 
relevant for the LIGO/Virgo observed events.  

The theoretical framework for the PBH formation that we shall follow next is based 
on the traditional Press-Schechter formalism \cite{Press:1973iz}.  
Large density perturbations can create overdense regions that 
may collapse to form black holes after the horizon reentry. 
We examine separately the two most interesting cosmological scenarios 
for the very early universe: the radiation (RD) and matter domination (MD) scenarios.

\subsubsection{Radiation-dominated era}
For a Gaussian distribution function of the primordial density  
perturbations and  for spherically symmetric regions, the 
mass fraction of PBHs at formation is 
\begin{equation} \label{brad}
\beta_\text{}(M)=\int_{\delta_c}d\delta\frac{1}{\sqrt{2\pi\sigma^2(k)}}e^{-\frac{\delta^2}{2\sigma^2(k)}}\,
\simeq \, \frac{1}{2}\text{erfc}\left(\frac{\delta_c}{\sqrt{2}\sigma(k)}\right) 
\, \simeq \,  \frac{1}{\sqrt{2\pi}} \frac{\sigma(k)}{\delta_c} e^{-\frac{\delta^2_c}{2\sigma^2(k)}}  \,.
\end{equation}
The parameter $\delta_c$ is the threshold density perturbation and erfc$(x)$ is the complementary error function.  For $\delta>\delta_c$ density perturbations overcome internal pressure and collapse. The $\beta$ parameter  can be regarded as the probability that 
the density contrast is larger than $\delta_c$.
The PBH abundance is exponentially sensitive to the threshold $\delta_c$.
Different values for $\delta_c$ are quoted in the literature, 
see e.g. \cite{Carr:1975qj, Niemeyer:1997mt,  Shibata:1999zs, Musco:2008hv,  Musco:2012au, Harada:2013epa, Germani:2018jgr, Byrnes:2018clq}, so that 
its precise value seems to be rather uncertain.  
In the comoving gauge, ref. \cite{Harada:2013epa} finds 
$\delta_c=0.41$ for $w=1/3$.
Numerical simulations demonstrate that there is no unique value 
for the threshold, because it depends on the density profile.

In the comoving gauge, assuming a nearly scale-invariant curvature power 
spectrum for a few e-folds around horizon crossing,
the curvature perturbation  ${R}$ can be related to the 
density perturbation $\delta$
as $\delta(k, t) = 2(1+w)/(5+3w)\left({k}/{aH} \right)^2 {R}(k, t)$.
The variance of the density perturbations $\sigma(k)$, smoothed on a scale $k$ in the radiation-dominated era, is given by \cite{Young:2014ana}

\begin{equation}
\sigma^2(k)= \left( \frac{4}{9} \right)^2  \int \frac{dq}{q}W^2(qk^{-1})(qk^{-1})^4{\Delta^2_R}(q)\,,
\end{equation}
where ${\Delta^2_R}(q)$ is the power spectrum of the curvature perturbations,
usually calculated numerically. 
Here $W(z)$ represents the Fourier transform of the Gaussian window function. 
In order to estimate the mass spectrum of the PBHs,  
the horizon scale  at the time of reentry of the perturbation mode $k$ has to be related to the mass of formed PBHs. 
 During the radiation era, the wavenumber scales as $k_\text{} \propto g^{1/2}_* g^{-2/3}_s  S^{2/3} a^{-1}$ and the 
 Hubble horizon as $H \propto g^{1/2}_* g^{-2/3}_s  S^{2/3} a^{-2}$, where $S$ denotes the entropy, and $g_*$, $g_s$ count the total number of the effectively massless degrees of freedom for the energy and entropy densities respectively. 
 Assuming conservation of the entropy between the reentry moment and the epoch of radiation-matter equality,
   the relation between the PBH mass $M$ and the comoving wavenumber $k$ is given by 
\begin{align}\label{mas}
 M(k)=\gamma \rho \frac{4\pi H(k)^{-3}}{3} \Big|_{k=a H}
\simeq 2.4 \times 10^{-16} M_{\odot}\Big(\frac{\gamma}{0.2}\Big)\left(\frac{g_*(T)}{106.75}\right)^{-\frac{1}{6}}\left(\frac{k}{10^{14}\,\text{Mpc}^{-1}}\right)^{-2},
\end{align}
where we took the effective degrees of freedom $g_*$ and $g_s$ approximately equal.  
The factor $\gamma$ gives the fraction of the horizon mass $M_H$ 
that collapses to form PBHs. Its value depends on the details of the gravitational collapse and an analytical estimation  \cite{Carr:1975qj}  gives $\gamma=0.2$. 
The present ratio of the abundance of PBHs  with mass $M$ over the total dark matter (DM) abundance,  $f_\text{PBH}(M)\equiv {\Omega_{\text{PBH}}(M)}/{\Omega_{\text{DM}}}$,  can be expressed as 
 \begin{equation}
f_\text{PBH} (M) \equiv \frac{\Omega_\text{PBH}}{\Omega_\text{DM}}\, =\,\left(\frac{\beta_\text{}(M)}{3.3 \times 10^{-14}}\right) \, 
\left(   \frac{\Omega_{\text{DM}}h^2}{0.12}   \right)^{-1}  
 \Big(\frac{\gamma_\text{}}{0.2}\Big)^{\frac{3}{2}} 
\left(\frac{g_*}{106.75}\right)^{-\frac{1}{4}}  
\left(\frac{M}{10^{-12} M_\odot}\right)^{-1/2}\,.
\end{equation}

The abundance of  PBHs produced during RD can be significant if the scalar spectrum is amplified by roughly 7 orders of magnitude.
In our single field models, described by the the $\alpha$-attractors potential (\ref{Vall}), such an enhancement is achieved if the potential involves several steps or a step and an inflection point. 
 In fig.   \ref{figfPBH} we plot the PBH fractional abundance for a potential with a step and and inflection point, 
 for the parameter values listed in section \ref{Secinfmodels}. 
 The scalar power spectrum of this model is depicted in 
 fig. \ref{figPRandGW}. For the estimation of the PBH abundance we 
 assumed a threshold value $\delta_c=0.45$ \cite{Musco:2012au}.  
 We see that, although the scalar power  spectrum is characterized by an oscillatory pattern around the peak of the PBH abundance, 
 it is predominantly monochromatic. 
 However, the induced GW spectrum is much more informative, 
 as we will discuss in the following.

\subsubsection{Matter-dominated era}
 PBHs might also form in the matter-dominated era (MD). 
In the absence of pressure, even minute perturbations will evolve
 and deviations from spherical configurations play an essential role.
Refs. \cite{Khlopov:1980mg, Polnarev:1986bi, Harada:2016mhb} examined the PBH production in a matter-dominated universe and considered the non-spherical effects during gravitational collapse.
The PBH production rate  $\beta$ tends to be proportional to the fifth power of
the variance $\sigma$  \cite{Harada:2016mhb}:
\begin{equation} \label{bmat}
\beta(\sigma)\, = \, 0.056 \, \sigma^5\,.
\end{equation}  
This expression has been derived with semi-analytical calculations and applies to
  $0.005 \lesssim \sigma \lesssim 0.2$, whereas  for $\sigma \lesssim 0.005$  the PBH production rate is modified if there is significant angular momentum 
  in the collapsing region \cite{Harada:2017fjm}.
 The PBH fractional abundance is 
\begin{align} 
f_\text{PBH}  \simeq \, 1.3 \times 10^{9} \, \gamma_\text{}\,  \beta\,  \frac{T_\text{rh}}{\text{GeV}}\, ,
\label{ratio}
\end{align}
with $T_\text{rh}$ is the reheating temperature.

\begin{figure} 
   \centering
   \includegraphics[width=.9 \linewidth]{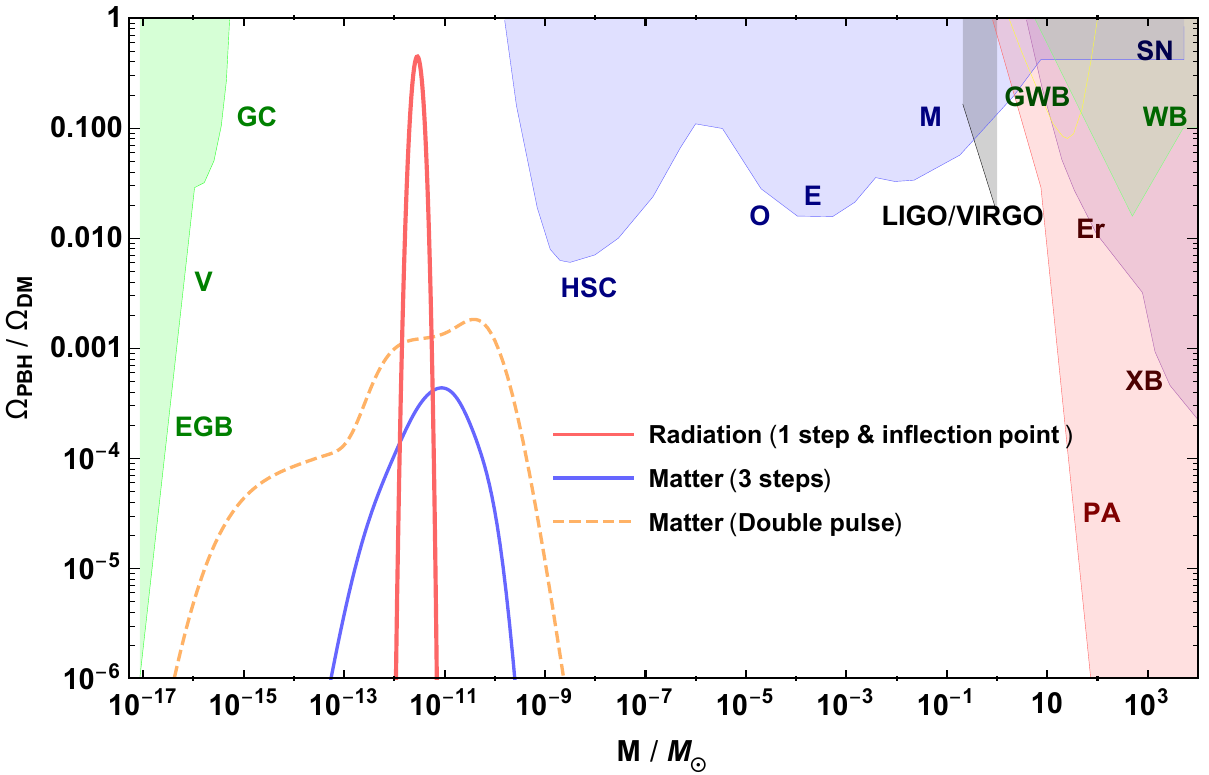}
  \caption{\label{figfPBH}~~  
 The red curve depicts the abundance of PBHs produced by an inflationary model with a step and an inflection point during the RD era. 
 The blue  curve depicts the abundance of PBHs produced by an 
 inflationary model with three steps
 during the MD era. The dashed curve is the PBH abundance 
produced during the MD era
 for the double-``pulse" model (termed Ex. 3) of fig. \ref{spectrum3}, in which
 a negative-friction ``pulse" is followed by a positive-friction ``pulse".
   }
 \end{figure}

There are two very interesting 
implications of PBH production during the MD era. 
Firstly, the PBH abundance is found to be 
larger compared to RD for a given amplitude of the  curvature power spectrum. 
Inflationary potentials with steps, which enhance the curvature power spectrum by 4
or 5 orders of magnitude,
can  have an observational effect by generating 
a significant cosmological PBH abundance.
Secondly, the PBH production during the MD era yields a PBH mass spectrum that is not predominantly monochromatic. It has a distribution over a few orders of the PBH mass values, which might reveal a non-trivial shape for the underlying power spectrum  of the primordial density perturbations. 
Although the specific inflationary models that we examine here do not have a
very strong effect on the
PBH mass spectrum,
inflationary models with steps can in principle produce mild modulations in the distribution  of PBHs. 
The blue curve  in fig. \ref{figfPBH} depicts the PBH abundance produced by an inflationary potential given by eq. (\ref{Vall}) with three steps, for 
amplitude $\Delta^2_R \sim 10^{-4}$ 
and $T_\text{rh}\sim 10^3$ GeV.
In the same figure, the dashed  curve depicts the PBH abundance produced by the 
double-``pulse" model of fig. \ref{spectrum3}
for $\Delta^2_R \sim 10^{-3}$ and $T_\text{rh}\sim 1$ GeV.
The spectrum is sufficiently wide and oscillating in order to 
have an observational impact on the PBH mass distribution.

 \subsection{Induced gravitational waves} \label{gravitationalwaves}

Primordial density perturbations that seed PBHs also produce stochastic GWs 
through the mode-mode coupling of the density perturbations beyond the linear order in 
the perturbative expansion. 
 The GW production takes place mainly at the time when the perturbations reenter the Hubble horizon. If density perturbations enter during the RD era, 
 the stochastic spectrum of second order GWs can be computed following cosmological perturbation theory \cite{Matarrese:1992rp,Matarrese:1993zf,Matarrese:1997ay, Mollerach:2003nq,  Noh:2004bc, Carbone:2004iv, Nakamura:2004rm, Baumann:2007zm, Ananda:2006af, Assadullahi:2009nf}. The same density perturbations will also produce PBHs with abundance proportional to $\beta$ given by eq. (\ref{brad}). On the other hand, if perturbations enter deep in an early MD era, 
 a different analysis has to be followed in order to find the GW energy density spectrum \cite{Jedamzik:2010hq, Dalianis:2020gup}. The corresponding PBH abundance will now be proportional to $\beta$ given by eq. (\ref{bmat}). 
 In the following we will consider GW production only during the RD era, leaving the study of the early MD era scenario for future work. 
   We will also assume that curvature perturbations are described by Gaussian statistics\footnote{Non-Gaussian statistics may also  generate modulations in the GW energy density spectrum \cite{Cai:2018dig}.}.

 The spectrum of the induced GWs is sourced and shaped by the 
curvature perturbations.
 In section \ref{notes} we found that inflationary potentials
 with steps generate a distinct oscillatory profile for the 
 curvature power spectrum. We expect that this profile  is transmitted to GWs. 
 In the following subsections we will further elaborate on 
 the modulations of the amplitude in the GW energy density spectrum, which will 
 be found to display a multiple peak structure. 
 We will show in particular that the  amplitude and the frequency of the peaks in the GW spectrum  are determined by the position and the number of the steps in the inflaton potential. 
 The GW spectrum inherits the pattern characteristics of the curvature power spectrum and, hence, serves as a portal to the inflationary dynamics.
 
 Different GW experiments are sensitive to different frequency bands.
Curvature power spectra with a prominent peak at the horizon mass range 
$10^{-15}-10^{-10} M_\odot$ 
generate induced GWs at the frequency band $1-10^{-4}$ Hz, and
can be tested by space-based interferometers like LISA \cite{Audley:2017drz}, 
scheduled to operate in the following decade.

\subsubsection{The formalism of induced GWs}

 GWs are described by the tensor perturbation $h_{ij}$ in the FRW spacetime
 \begin{align}
 ds^2=a^2(\tau)\left[-(1+2\phi)d\tau^2 +\left((1-2\psi)\delta_{ij} +\frac{1}{2}h_{ij}   \right)dx^idx^j \right],
 \end{align}
 where $\phi$ and $\psi$ are scalar perturbations and vector perturbations are
 neglected.
 In the absence of anisotropic stress, which is a good approximation for our 
 purposes, we have $\phi=\psi$. 
 The Fourier components of the tensor modes are
 \begin{equation}
 h_{ij}(\tau, \textbf{x})\, = \,\sum_\lambda  \int \frac{d^3 k}{(2\pi)^{3/2}} h_\lambda(\tau, \textbf{k}) e^{(\lambda)}_{ij} (\textbf{k}) e^{i\textbf{k}\textbf{x}}
 \end{equation}
where   $e^{(\lambda)}_{ij}$, with $\lambda=+, \times$, are polarization tensors.
 Through the definition of the dimensionless power spectrum 
 \begin{equation} \label{Pdef}
 \left\langle h_\lambda(\tau, \textbf{k})h_\lambda(\tau, \textbf{k}') \right\rangle =\delta_{\lambda \lambda'} \delta^3(\textbf{k}+\textbf{k}') \frac{2\pi^2}{k^3} {\cal P}_h(\tau, k)
 \end{equation}
 we have
 \begin{equation}
 \rho_\text{GW}(\tau, k)= \frac{M^2_\text{Pl}}{8} \, \frac{k^2}{a^2}\,  \overline{{\cal P}_h(\tau, k)}\,.
 \end{equation}
 The evolution of $h_{ij}$ is obtained by expanding the Einstein equations. At second order in scalar perturbations, the  
  equation of motion for the Fourier components of the tensor perturbations is 
 \begin{equation} \label{tensorsEq}
 h''_\lambda+2{\cal H} h'_\lambda+k^2 h_\lambda \, = \, 4 S_\lambda(\tau,\textbf{k})\, ,
 \end{equation}
 where $ S_\lambda(\tau,\textbf{k})$ is a source that consists of products of scalar perturbations:
 \begin{equation}
  S_\lambda(\tau,\textbf{k})=\int \frac{d^3 k}{(2\pi)^{3/2}} e^{ij}(\textbf{k})q_i \,q_j \left( 2\phi_\textbf{q}\phi_{\textbf{k}-\textbf{q}}+\frac{4}{3(1+w)}({\cal H}^{-1} \phi_\textbf{q}+\phi_\textbf{q})({\cal H}^{-1} \phi_{\textbf{k}-\textbf{q}}+\phi_{\textbf{k}-\textbf{q}}) \right)\,.
 \end{equation}
 The evolution of  $\phi_\textbf{k}$  
 is given in terms of the scalar transfer function. For radiation domination, we have
 $\phi_\textbf{k}(\tau)= \phi(x)\ \Phi_{\textbf{k}}$, with
 \begin{equation}
 \phi(x)=\frac{9}{x^2}\left( \frac{\sin(x/\sqrt{3})}{x/\sqrt{3}} -\cos(x/\sqrt{3})\right), 
 \end{equation} 
 where $x=k\tau$,  and $\Phi_\textbf{k}$ is the primordial value,
 related to the curvature perturbation as
 \begin{equation} \label{prim_phi}
 \left\langle \Phi_\textbf{k} \Phi_{\textbf{k}'} \right\rangle= \delta^3(\textbf{k}+{\textbf{k}'}) \frac{2\pi^2}{k^3} \left( \frac{3+3w}{5+3w}\right)^2  \Delta_R^2(\tau, k).
 \end{equation}
 
The solution of eq. (\ref{tensorsEq}) reads
 \begin{equation}
 h_\lambda(\tau, \textbf{k})=\frac{1}{a(\tau)} \int_0^\tau G_k(\tau, \bar{\tau}) a(\bar{\tau}) S_\lambda(\bar{\tau}, \textbf{k}) d\bar{\tau}.
 \end{equation}
 where $G_k(\tau, \bar{\tau})$ is the the Green function for eq. (\ref{tensorsEq}).
The power spectrum of induced GWs is expressed in a compact form as a double integral involving the power spectrum of the curvature perturbations \cite{Kohri:2018awv}
 \begin{equation}
 \label{tensorPSD2}
     \overline{\mathcal{P}_{h}(\tau,k)}=\int_{0}^{\infty} dt \int_{-1}^{1} ds\ \mathcal{T}(s,t,\tau,k)\ \Delta_R^2\left(\frac{t+s+1}{2}k\right)\ \Delta_R^2\left(\frac{t-s+1}{2} k \right)\,.
 \end{equation}
 The overline denotes the oscillation average. The $t$ and $s$ variables are 
 defined as  $t=u+v-1$, $s=u-v$, where  $ v={q}/{k}$, $u=|\textbf{k}-\textbf{q}|/{k}$.
 The integral kernel ${\cal T}$  is given by the expression
 \begin{align}
 \label{radiation_TTF}
     \lim_{x\to\infty} x^2\ \mathcal{T}(s,t,x)& = \ 2
 \left(\frac{t(2+t)(s^2-1)}{(1-s+t)(1+s+t)} \right)^2      
 \frac{288(-5+s^2+t(2+t))^2}{(1-s+t)^6(1+s+t)^6}   \\
 & \times \left\{ \frac{\pi^2}{4}\left(-5+s^2+t(2+t)\right)^2\Theta 
 \left( t-(\sqrt{3}-1\right)+ \right.   \nonumber \\
 &
 \left. \left(-(t-s+1)(t+s+1)+\frac12(-5+s^2+t(2+t)) \log \left|\frac{-2+t(2+t)}{3-s^2} \right|\right)^2 \right\} \,,  \nonumber
 \end{align}
 where $\Theta$ is the Heaviside step function.
The fraction of the GW energy density per logarithmic wavenumber interval is 
 \begin{align} \label{OmegaIGWc}
 \Omega_\text{GW}(\tau, k)
 =\frac{1}{\rho_\text{tot}(\tau)} \frac{d \rho_\text{GW}(\tau, k)}{d\ln k} =\frac{1}{24}\left(\frac{k}{a(\tau) H(\tau)} \right)^2 \overline{{\cal P}_h(\tau, k)} \,.
 \end{align} 
 At a certain time $t_\text{c}$ 
 the production of induced GWs ceases, while their propagation becomes free. 
 In a RD background the energy density parameter $\Omega_{\rm GW}$ remains constant.
 Its value at the current time $t_0$ is given by eq.~(\ref{OmegaIGWc}) 
 times the current radiation density parameter, $\Omega_{\gamma,0}h^2=4.2\times10^{-5}$, modulo changes in the number of the relativistic degrees of freedom $g_*$ in the radiation fluid:
 \begin{align}
 \label{IGWtoday}    
     \Omega_{\textrm{GW}}(t_0,f)h^2=0.39\times\left({g_{\ast}\over106.75}\right)^{-{1/3}}\ \Omega_{\gamma,0} h^2\times\Omega_{\textrm{GW}}(t_\text{c},f).
 \end{align}
 The total energy density parameter of induced GWs is obtained by integrating the GW energy density spectrum over the entire frequency interval.

\begin{figure} 
   \centering
\includegraphics[width=0.49\linewidth]{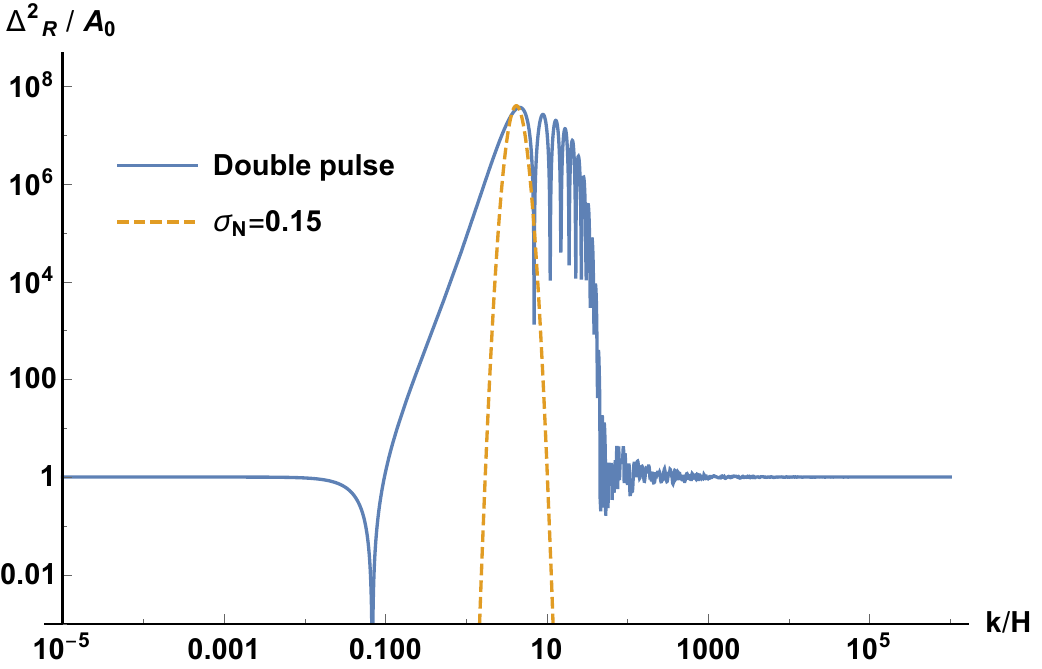}
\includegraphics[width=0.49\linewidth]{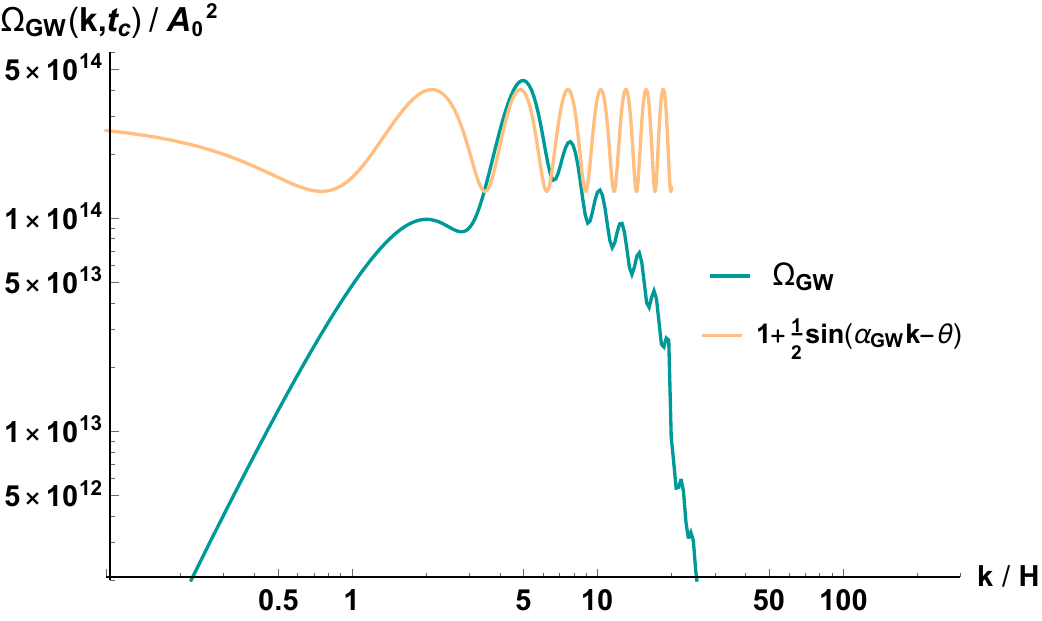}
\includegraphics[width=0.49\linewidth]{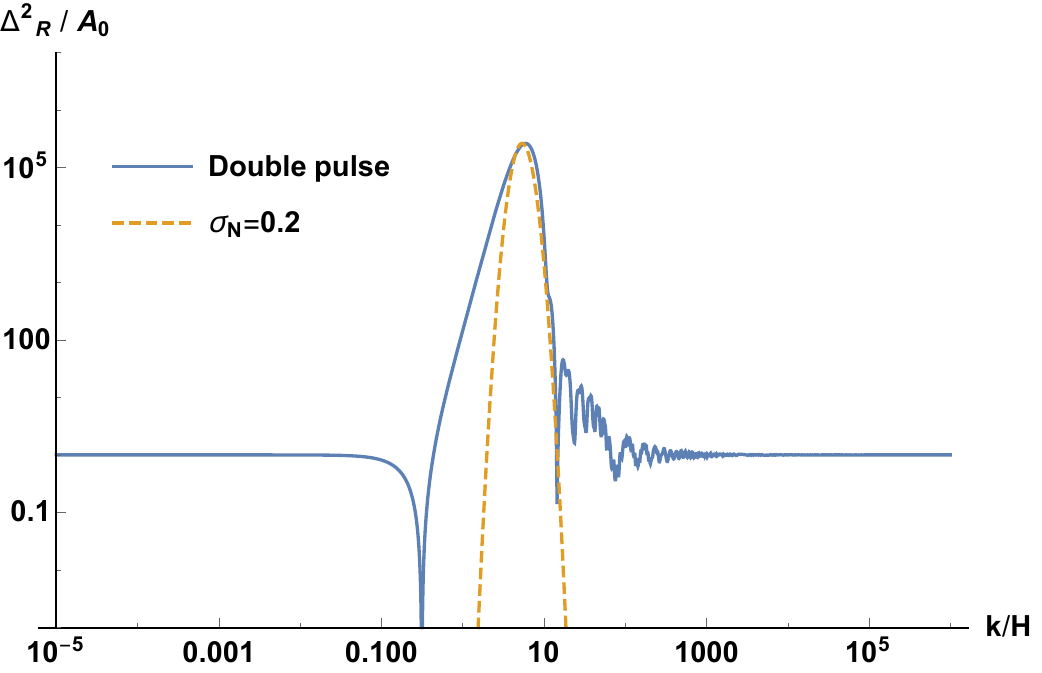}
\includegraphics[width=0.49\linewidth]{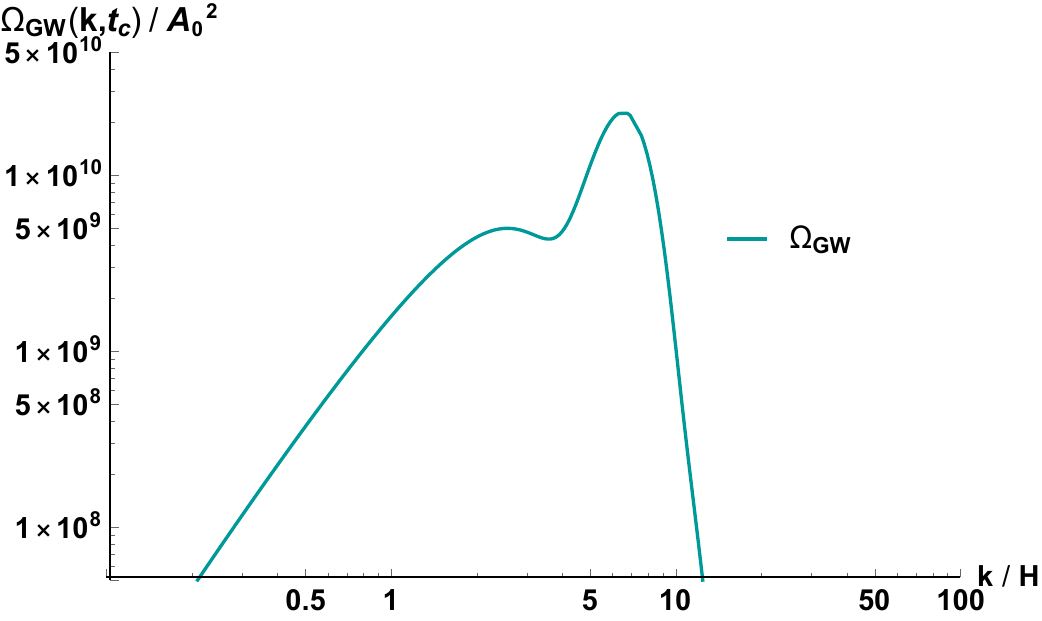}
\includegraphics[width=0.49\linewidth]{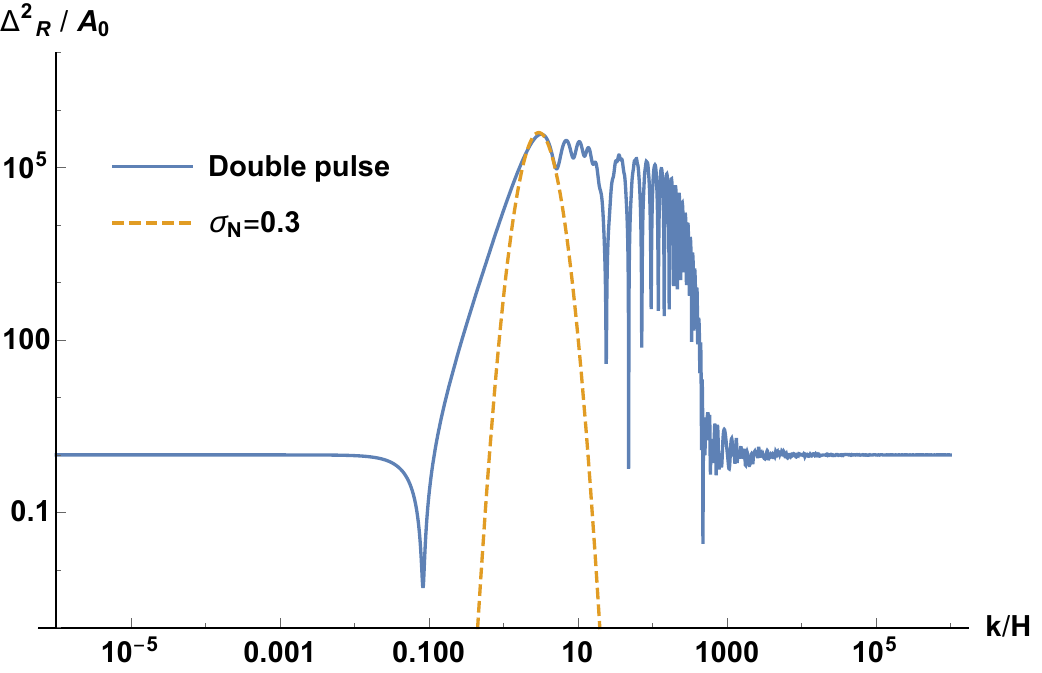}
\includegraphics[width=0.49\linewidth]{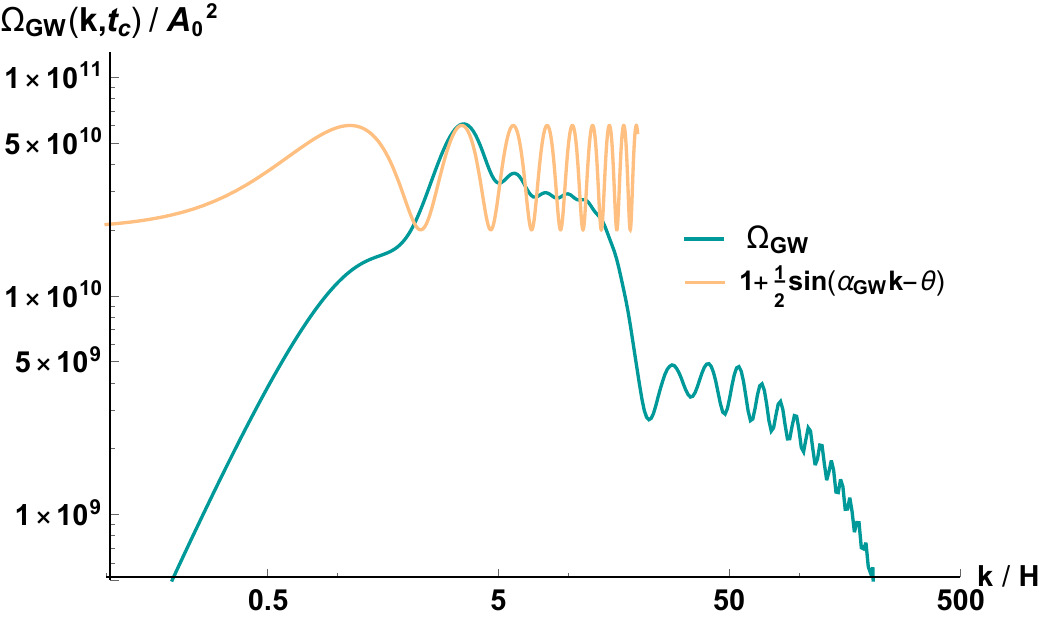}
  \caption{\label{figPRexGW}~~  
Left: The scalar power spectrum produced by a double square ``pulse", normalized to
 the CMB measured amplitude.
 The dashed line is a lognormal power spectrum with width $\sigma_\text{N}$. 
 Right: 
The induced GW spectrum, along with a fitting harmonic function. 
Each row, from top to bottom, corresponds to the 
``pulse" producing the spectrum of each of 
figs. \ref{spectrum1}, \ref{spectrum2}, \ref{spectrum3}, 
respectively. 
}
 \end{figure}

\begin{figure} 
   \centering
   \includegraphics[width=.45 \linewidth]{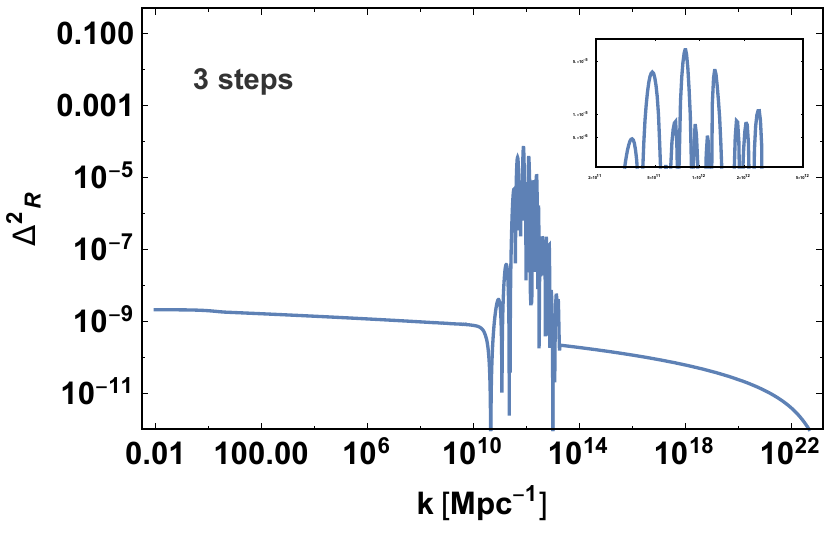}
   \includegraphics[width=.45\linewidth]{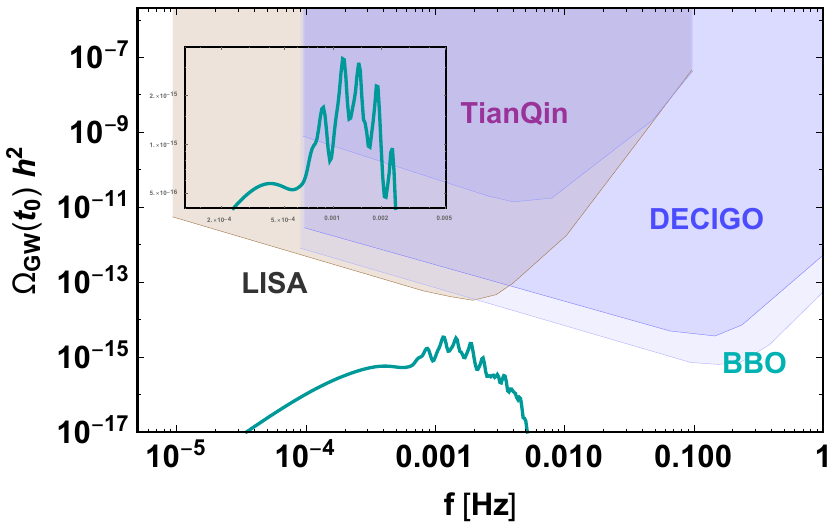}\\
      \includegraphics[width=.45 \linewidth]{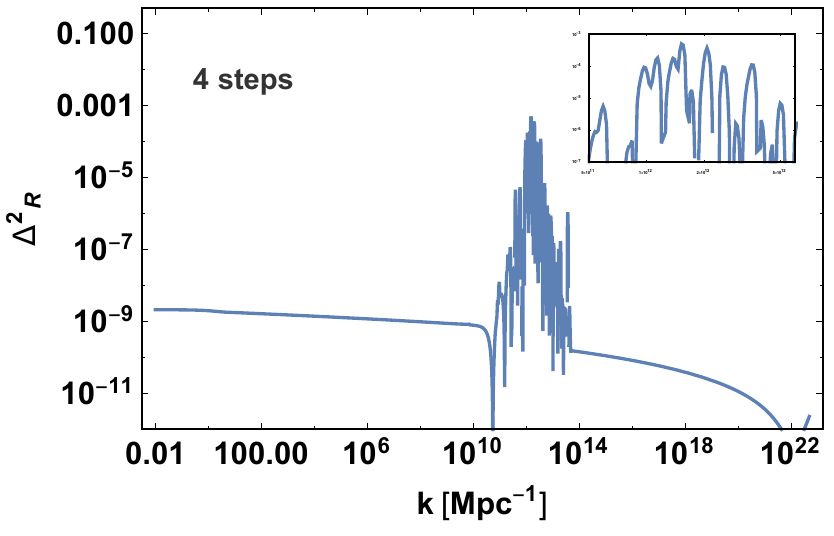}
   \includegraphics[width=.45\linewidth]{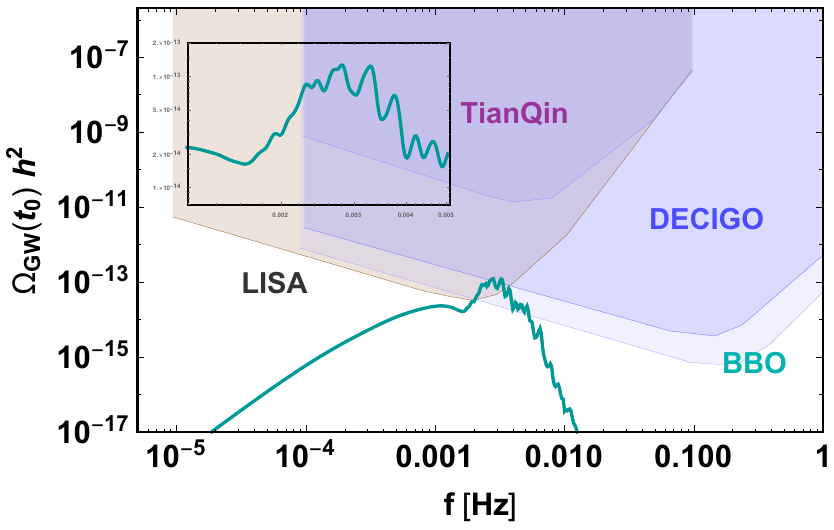}\\
     \includegraphics[width=.45 \linewidth]{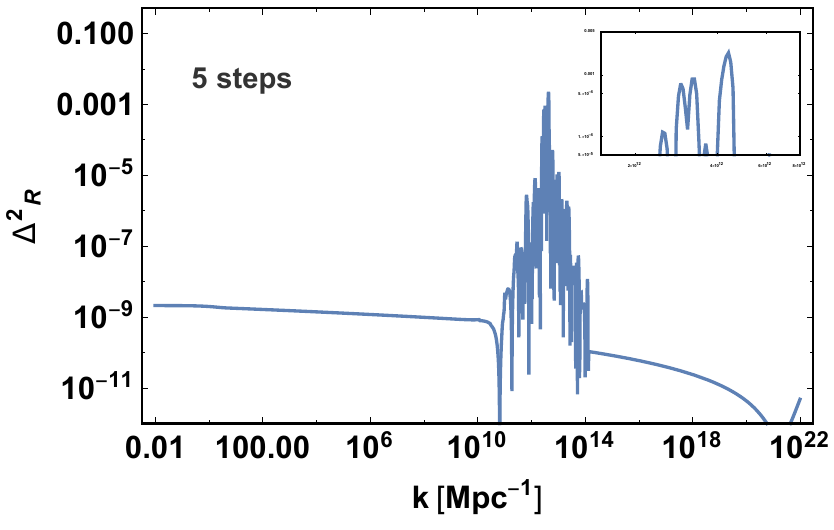}
   \includegraphics[width=.45\linewidth]{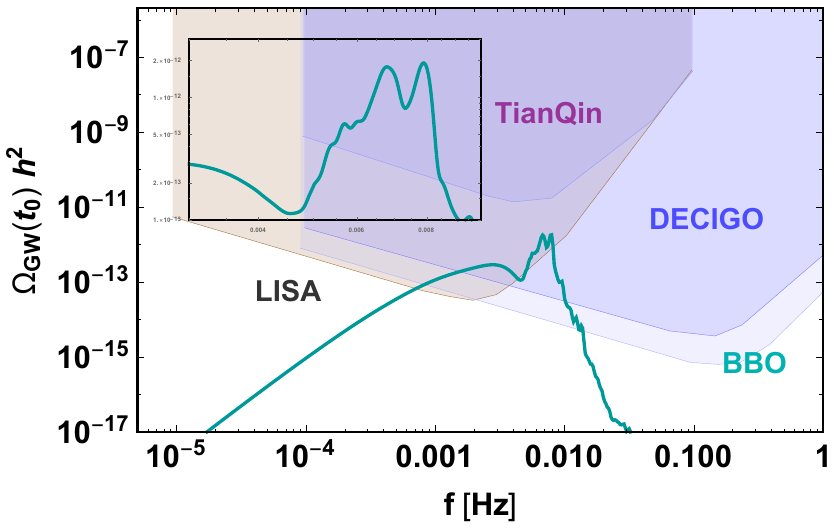}\\
   \includegraphics[width=.45 \linewidth]{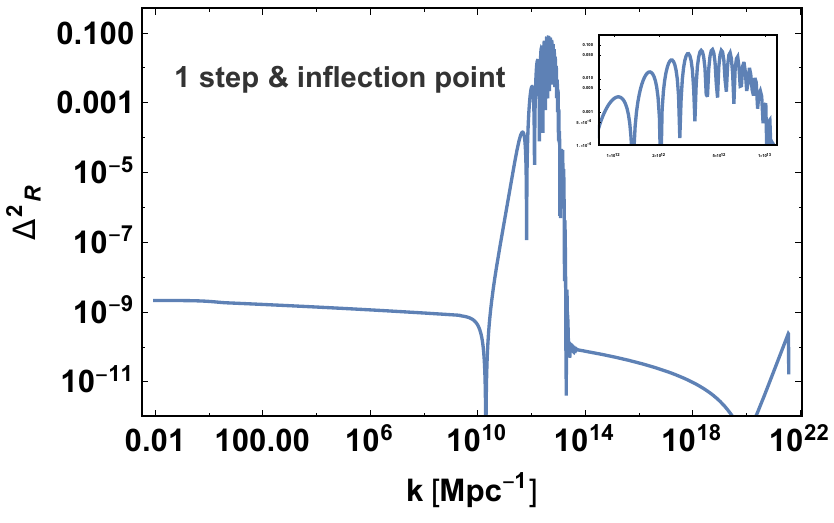}
   \includegraphics[width=.45\linewidth]{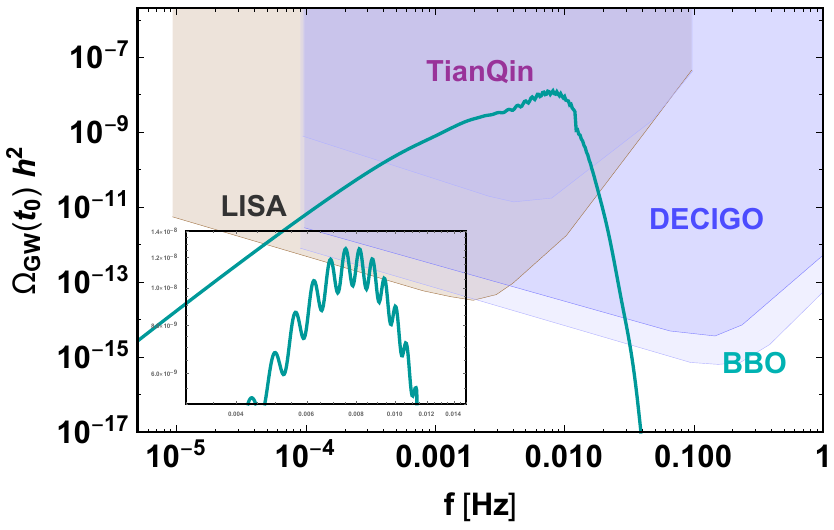}
  \caption{\label{figPRandGW}~~  
Left panels:  
The curvature power spectra produced by 
inflationary models with potential given by eq. (\ref{Vall})
and depicted in fig. \ref{FigVstepsCollection}, 
for parameter values given in the text. 
Right panels: The generated GW density parameter produced by each  inflationary model.
A zoom-in plot of the peak region has been included in each panel.
Note that the last  row of panels corresponds to an inflationary model that features both a step and an inflection point, depicted in fig. \ref{FigVstepInflection}.
   }
 \end{figure}

 \subsection{Oscillations in the power spectrum of the induced GWs from potentials with steps}

We start the discussion of the pattern 
of induced GWs 
produced in inflationary models with sharp features by looking at the spectrum characteristics of analytically calculable models, such as those
depicted in figs. \ref{spectrum1}-\ref{spectrum3}.
In section \ref{notes} we performed a semi-analytic calculation of the curvature power spectrum by modeling the function  $f(N)$ of eq. (\ref{AN}), 
which captures the dynamics of the inflaton field beyond the slow-roll regime, 
through  a sequence  of  square ``pulses".
The amplitude of the produced curvature power spectrum is enhanced by 
the factor $|C_m|^2$ of eq. (\ref{enhancement}), while it also
displays oscillatory patterns with characteristic periods 
$\delta k /H \sim e^N \pi$ in $k$-space, where $N$ is the number of efoldings at
which the function $f(N)$ varies strongly. (See the discussion below eq. (\ref{enhancement}).)
Roughly the same characteristic frequency is observed in the GW spectra.  
In fig. \ref{figPRexGW} the GW  spectra
 for the three examples studied in  section \ref{form} are plotted. 
We also plot a harmonic function $\Omega_\text{GWmax}(2+ \sin(\alpha_\text{GW}\, k -\theta))/3$ 
  that  highlights the periodic change of the  GW amplitude around the peak through
  a fit of the $k$-space period 
 $\delta k_{\text{GW}}  \sim  {2\pi}/{\alpha_\text{GW}}$.
 We find $\alpha_\text{GW} ={\cal O}(1)/H$, corresponding to the
 smallest period of the oscillations in the curvature spectrum 
 $\delta k/H \sim {\cal O}(1)\pi$.
It must be noted that larger periods of size $\delta k /H = {\cal O}(10)\, \pi$ 
appearing in $\Delta^2_R$ can also be 
discerned in the modulations of the amplitude of the GW spectrum at the corresponding scales. However, they are less visible as they extend to regions in $k$-space 
far from the peak.


The short-period modulations of order $\pi$ 
imply that the peaks in the curvature spectrum are narrow.
 In fig. \ref{figPRexGW} a lognormal distribution with a certain 
 width $\sigma_\text{N}< 1$  is plotted  together with $\Delta^2_R$. 
It elucidates the prominent two-peak structure induced in the GW spectrum \cite{Ananda:2006af, Saito:2008jc, Pi:2020otn}, which appears when 
the main peak of the curvature spectrum is sufficiently narrow. 
In our first example, for a square pulse starting at $N=0$, corresponding 
to a smallest period of oscillations $\delta k/H\simeq \pi$, 
the peak of $\Delta^2_R$ is found at a wavenumber $k_\text{p}/H\simeq 4.5$, 
comparable to $\pi$. 
The $k$-range of the fitting lognormal distribution is determined by
requiring that 
$\delta k \sim k_\text{p} (e^{\sigma_\text{N}}-1) $, which implies that  
$\sigma_\text{N}< 1$.
As a result, and in agreement with the analysis of ref. \cite{Pi:2020otn}, 
the GW spectrum is found to  feature a major, relatively sharp peak
at $(2/\sqrt{3})k_\text{p}/H$. Additionally, in the low-$k$ side there 
is a relatively flat local maximum, at a wavenumber near $k_\text{p}/e$.
This characteristic two-peak structure is evident in all three examples we 
studied. In the second example in particular, in which $\Delta^2_R$ is dominated by a single peak because the smallness of the positive ``pulse"
confines the oscillatory patterns within the high-$k$ part of the spectrum, the two-peak structure is practically the only observable feature.

Let us now turn to the inflationary models with step-like features 
described by eq. (\ref{Vall}). 
In these models, the effective-friction function $f(N)$ of eq. (\ref{AN}), 
depicted in fig. \ref{FigVstepsCollection}, 
has a $2n$-``pulse" structure for $n$ steps, 
with each positive ``pulse" followed by a negative one.
In the semi-analytically calculable models that we studied before,  
the power spectrum was normalized such that the step features started at $N=0$. 
We observed an enhancement by a factor 
$|C_m|^2$, together with oscillations of period $\delta k/H\sim e^N \pi$.  
In the inflationary models of eq. (\ref{Vall}) we find numerically similar patterns. 
The curvature spectrum $\Delta^2_R$ is to a good approximation  
enhanced by $|C_m|^2$, with a main 
peak at a wavenumber $k_\text{p}$ characteristic of the step position in terms of
the number of efoldings $N$, which
are now counted from the exit of the CMB scale. 
Oscillations are also 
produced with an approximate period  $\delta k\sim k_p$.  

In fig. \ref{figPRandGW}  we plot four curvature power spectra together 
with the GW density spectra that they induce.
For the three-step model, 
the curvature  spectrum displays strong modulations and 
one can read off an oscillatory pattern with period 
$\delta k \simeq 2.5\times 10^{11} \,\text{Mpc}^{-1}$.
The spectrum $\Delta^2_R$ has three prominent peaks at comparable 
wavenumbers 
$k_{\text{p}1}$,  $k_{\text{p}2}$,  $k_{\text{p}3}$. 
Each peak is well described by a narrow lognormal distribution in the $k$-range
with width $\sigma_\text{N}< 1$.  
The combination of these three peaks induces a characteristic five-peak structure 
in the GW spectrum \cite{Cai:2019amo}, along with the rather flat local maximum at lower $k$,
as can be seen in the first row of panels in fig. \ref{figPRandGW}. 
The first sharp peak in the GW spectrum is located at the value 
$k_\text{GW,1}=2 k_{\text{p}1}/\sqrt{3}$, the second at  $k_\text{GW,2}=(k_{\text{p}1}+k_{\text{p}2})/\sqrt{3}$, the third at $k_\text{GW,3}=2 k_{\text{p}2}/\sqrt{3}$, the fourth at $k_\text{GW,4}=(k_{\text{p}2}+k_{\text{p}3})/\sqrt{3}$ and the fifth at $k_\text{GW,5}=2 k_{\text{p}3}/\sqrt{3}$.
The rather flat local maximum at lower $k$ is located near $k_{\text{p}2}/e$, where $k_{\text{p}2}$ is the wavenumber of the highest peak.
These values  can be seen in 
fig.  \ref{figPRandGW} in the frequency spectrum and 
in Hz units through the conversion $f=k/(2\pi)$, 
where $\text{Mpc}^{-1}\simeq 0.97 \times 10^{-14}$ Hz.
Similar conclusions can be drawn for the next two inflationary 
models that feature four and five steps, respectively.  

For the inflationary model that features both an inflection point and a step, 
the resulting spectra are quite different compared to the previously discussed models that involve only steps. The inflection point is responsible for the strong enhancement of the curvature power spectrum and its relatively wide peak. 
Indeed, the envelope function that outlines the peak can be fitted by a lognormal distribution 
 with width $\sigma_\text{N}\simeq 0.4$. Hence  the characteristic  two-peak structure 
 in the GW spectrum \cite{Pi:2020otn} is not very prominent here. 
The step-like feature has a minor contribution to the enhancement of the curvature spectrum, 
but it is the source of the oscillatory pattern around the peak with the characteristic period 
$\delta k \simeq 7\times 10^{-11} \text{Mpc}^{-1}$.
These oscillations are also transferred to the GW spectrum. 
As before, the GW oscillatory pattern is well described by a harmonic function and
reflects the pattern in the curvature spectrum.

It is important to emphasize that, even though it is not clearly visible 
in the log-plot,
the oscillations near the peak are substantial:
The $\Omega_\text{GW}$ spectrum displays variations in its
amplitude that are $25\%$ of its maximal value or larger. 
Such modulations in the amplitude are likely to be detectable 
by the near-future space interferometers.

\section{Conclusions} \label{conclusions}

In this paper we studied single-field inflationary models with sharp, step-like features in the potential. 
The evolution of the inflaton through such features leads to the 
violation of the slow-roll conditions, and in some cases even to the temporary 
interruption of inflation.
The striking consequences of the transition through a generic step-like feature 
are the enhancement of the power spectrum of the curvature perturbations at certain
scales by several orders of magnitude and the production of
distinctive oscillatory patterns.
We studied analytically and numerically the inflationary dynamics and we 
derived the expressions that describe quantitatively the size of the 
enhancement, as well as the profile of the oscillations.
It is interesting that these features impact the power spectrum 
in a distinctive and predictive way that reflects their properties: 
the amplification and the oscillatory pattern are shaped by 
the position, number and steepness of the features.

Our analysis has revealed the origin of the oscillations. They are 
generated through the detuning of the phase difference between the
real and imaginary parts of the curvature perturbation, which evolves
according to the Mukahnov-Sasaki equation (\ref{eomv}).
When the inflaton moves through a step in the potential, the 
background evolution deviates strongly from the standard slow-roll
for a small number of efoldings, in a way that 
the real and imaginary parts of the solution are detuned. The detuning
results in time-dependent oscillations of the amplitude. When the
perturbations asymptotically freeze at superhorizon scales, an 
oscillatory pattern is induced on the wavenumber dependence of the
power spectrum. A detailed discussion of this point can be found
in ref. \cite{Kefala:2020xsx} and subsections \ref{toy}, \ref{analytexpr}, \ref{analytfN}.
It must be emphasized that oscillations in the spectrum are not a generic consequence of any
feature in the potential that violates slow roll. 
In contrast to a steep step, an inflection point
in general induces an enhanced, but smooth, power spectrum.

From the model-building perspective, a steep step can appear if the inflationary potential includes plateaus with different energy densities. 
A nearly constant potential energy density can be associated with 
underlying symmetries that are preserved in the plateau \cite{Kallosh:2014rga}.
A deformation of the symmetry results in energy splitting, 
so that a transition between different energy levels can be induced. 
Such a behavior can be captured by the framework of 
the inflationary models characterized as $\alpha$-attractors 
\cite{Kallosh:2013hoa, Ferrara}.

A strong motivation for analysing such models is that the  
induced tensor power spectrum inherits the oscillating profile of 
the primordial curvature spectrum. 
The combined pattern of an enhanced 
spectrum together with strong oscillations is potentially detectable 
by near  future space interferometers.
Through the detection of the GW spectrum, one can aim at inferring at least 
some basic feature of the 
inflationary potential, such as whether step-like transitions are present.
Motivated by this possibility, we examined in detail, 
numerically and analytically, 
the scalar and the induced tensor spectra and we identified correlations between 
them. The main characteristic property of both spectra, related to 
the transition through a step, is a series of peaks.
Through a more refined analysis of the spectrum, 
one can look for more detailed information, such as 
the number of the steps, their position and exact shape, and  whether there is, in 
addition to a step, an inflection point.
We explored this possibility by studying several analytical
examples, as well as inflationary models in the $\alpha$-attractor framework,
always imposing consistency with the 
contraints for the spectral index $n_s$ and the amplitude of the scalar spectrum 
arising from the CMB measurements.

The detection of GWs from inflationary models with sharp features may be 
accompanied by the presence of PBHs as a significant fraction of dark matter.
The enhancement of the power specrtum due to the presence of step-like 
features, though considerable, may  
be inefficient to trigger the production of a sizeable number of PBHs
if radiation dominates the energy density of the early universe. 
However, it can be sufficient to induce gravitational collapse 
processes and PBH production if the 
universe energy density is dominated by non-relativistic  matter. We examined the 
profile of the PBH mass spectrum produced either in a radiation or an early 
matter-dominated universe, 
looking for deviations from the common monochromatic profile. 
For the latter scenario we found that this is possible because of the 
multiple-peak structure of the curvature power spectrum.

It is important to note that
the oscillations near the peak of the GW spectrum have a scale 
comparable to that of its maximal value. 
Such modulations in the amplitude are likely to be detectable 
by the near future space interferometers, such as LISA.
This demonstrates that induced GWs can be used as a powerful 
tool for probing the inflationary potential. The detection of 
oscillatory patterns in the amplitude of the  GW spectrum  will be a 
strong indication for sharp features in the potential of single-field inflation.

\section*{Acknowledgments}

We would like to thank V. Spanos for useful discussions.
The work of I. Dalianis, G. Kodaxis, I. Stamou and N. Tetradis was supported by the Hellenic Foundation 
for Research and Innovation (H.F.R.I.) under the  “First Call for
H.F.R.I. Research Projects to support Faculty members and Researchers and the procurement of high-cost research equipment grant”  (Project Number: 824).


\end{document}